# A Study of Feshbach Resonances and the Unitary Limit in a Model of Strongly Correlated Nucleons

A.Z.Mekjian
Department of Physics and Astronomy, Rutgers University, Piscataway, NJ 08854

**Abstract**

*A model of strongly interacting and correlated hadrons is developed. The interaction used contains a long range attraction and short range repulsive hard core. Using this interaction and various limiting situations of it, a study of the effect of bound states and Feshbach resonances is given. The limiting situations are a pure square well interaction, a delta-shell potential and a pure hard core potential. The limit of a pure hard core potential are compared with the corresponding well known results for a spinless Bose and Fermi gas. The limit of many partial waves for a pure hard core interaction is also considered and result in expressions involving the hard core volume. This feature arises from a scaling relation similar to that for hard sphere scattering with diffractive corrections. Features that parallel the van der Waals gas and results based on a Skyrme interaction are discussed. The role of underlying isospin symmetries associated with the strong interaction of protons and neutrons in this two component model is investigated. Properties are studied with varying proton fraction. An analytic expression for the Beth Uhlenbeck continuum integral is developed which closely approximates exact results based on the potential model considered. Based on this analytic expression an analysis of features associated with a unitary limit is given. In the unitary limit of very large scattering length, the ratio of effective range $r_0$ to thermal wavelength $\lambda_T$ appears as a limiting scale in the interaction energy. Thermodynamic quantities such as the entropy and compressibility are also developed. The effective range corrections to the entropy vary as the cube of this ratio for low temperatures and are therefore considerably reduced compared to the corrections to the interaction energy which varies linearly with this ratio. Effective range corrections to the compressibility are also linear in the ratio.*

PACS 24.10.Pa, 21.65.Mn, 21.65.Cd, 67.85.Pq, 03.75.-b

Keywords: Two specie Fermi gases, Feshbach resonances and bound states, unitary limit, entropy and compressibility considerations

## I. Introduction

This paper is devoted to a study of properties of strongly correlated fermions and in particular to a two component hadronic system made of protons and neutrons. Extensions to a hadronic system which contains both neutrons and protons is a rich extension of the recent interest in one component Fermi systems such as in the field of condensed matter [1-4]. Even in the limit of neutron stars a small fraction of protons exist. Also, future FRIB (Facility for Rare Isotope Beams) experiments will study properties of nuclei with large neutron and proton excess. In a system of both protons and neutrons isospin symmetries and associated features come into play. The n-p system has a bound state in the spin $S=1$, angular momentum $L=0$ and isospin $I=0$ channel while the n-n, p-p and n-p systems in the $S=0$, $L=0$, $I=1$ channels have resonant like structures with a large scattering length. Thus, the nuclear system offers a unique situation with both bound and continuum resonant states present with an underlying isospin symmetry.

Of particular interest in the behavior of Fermi systems is in a limiting situation called the unitary limit. Bertsch formulated a challenge problem [5] based on this unitary limit. A unitary limit appears when the scattering length becomes infinite which occur when the bound state and resonant energies are set equal to zero. In atomic systems, Feshbach resonances are used as a tool for studying the unitary limit [1-4]. In particular tuning across a Feshbach resonance with an external magnetic field has been a very successful tool for the study from a Bose-Einstein condensate of tightly bound pairs to a BCS superfluid state. In atomic systems early studies of the unitary limit can be found in ref[6] at temperature $T=0$ and in ref [7]for $T \neq 0$. In nuclear physics, the unitary limit was initially studied for one-component systems of pure neutrons. Early work on the Berstch challenge problem was done by Barker [8] and latter by Heiselberg [9]. A Monte Carlo numerical study of the unitary limit of pure neutron matter is given in ref [10]. Analytic studies of pure neutron systems at zero temperature are given in the extensive work of Bulgac and collaborators [11-14]. The Bertsch problem was also addressed by Schwenk and Pethick[15]. Lattice calculations of the equation of state of neutron matter and nuclear matter have also appeared in the literature [16,17]. The work outlined in ref [16] showed a sharp transition from an uncorrelated Fermi gas to a cluster system characterized by a sharp peak in the incompressibility. Virial expansions of the equation of state are developed in ref[18-21] for nuclear matter and pure neutron matter. A recursive method approach to the nuclear canonical partition function in a cluster virial expansion has also been developed [22]. The resulting partition function showed a sharp first order liquid gas phase transition in which large cluster appear at a given temperature as the hadronic system is cooled below its boiling point [23-26].

The present manuscript involves further developments of a previous paper [27] which focused on the role of Feshbach resonances and limiting universal behavior in two component systems. Here a more detailed and in depth study of this earlier work is given as well as several new extensions of it. The previous work focused on $S-$wave two particle continuum correlations and the role of deuteron bound states. The present work investigates the role of higher partial waves. The contribution of many partial waves in the continuum correlations from a hard sphere gas results in expressions in the equation of state involving the hard sphere volume. This feature arises from a scaling law that parallels results for hard sphere scattering where an enhancement by a factor of two in the cross section arises from diffractive effects. The presence of the hard sphere volume in the equation of state is similar to that of a van der Waals gas.

This paper also includes a study of the entropy and compressibility of the system as well as other thermodynamic variables. The interaction used in this study is an attractive square well potential with a repulsive hard core. An attractive square well potential, a pure hard core gas and also a surface delta shell potential are special cases of the interaction used and results for these limiting cases are developed and compared. An analytic expression for the Beth Uhlenbeck continuum integral is shown to very accurately describe exact results based on the interaction potential model considered. Based on this analytic expression studies of features associated with a unitary limit are given. In the unitary limit of very large scattering length the ratio of the effective range $r_o$ to quantum thermal wavelength $\lambda_T$ appears as a limiting scale in the interaction energy. The effective range corrections in the unitary limit for the entropy are shown to be much smaller than for the energy because of a cancellation of two contributing terms. For the entropy these corrections vary as the cube of the ratio $r_o/\lambda_T$. For the interaction energy, Helmholtz free energy, chemical potential and compressibility these corrections vary linearly in $r_o/\lambda_T$. Results are also compared with pure hard sphere Bose and Fermi gases which also contain similar features. Studies are also done with a delta shell potential and compared with a square well potential. Similar thermodynamic results are found even though the phase shifts for these two potentials are quite different at high momentum. The Boltzmann factor in energy is shown to suppress the differences.

**II. Thermodynamics of Hadronic Systems.**
*General thermodynamic results*

This section summarize important general thermodynamic results for a high temperature nearly ideal gas. The details of the approach can be found in Ref [28,29,25]. Results for a one component system will be given first and latter subscripts will be added for a two component system of protons and neutrons. Furthermore, to illustrate results only symmetrization and antisymmetrization effects will be considered to begin with and then interaction effects will be included.

The canonical partition function can be written in terms of a set of variables called $x_k$ with $k = 1,2,3...A$ and $A$ is the size of the system. For an ideal gas, the $x_k$ [29, 30] are

$$x_k = (\pm 1)^{k+1} \frac{1}{k} \int \frac{4\pi p^2 V}{h^3} \exp[-\frac{p^2}{2mk_B T}]dp = \frac{(\pm 1)^{k+1}}{k^{5/2}} \frac{V}{\lambda_T^3} \tag{1}$$

with $\lambda_T^3 = h^3/(2\pi m k_B T)^{3/2}$. The $x_k$ are the cycle class weights coming from anti-symmetrization of fermions (the $(-1)^{k+1}$ term) or symmetrization of bosons (the $(+1)^{k+1}$ term) for a non interacting gas of either fermions or bosons. The canonical partition function $Z_A = Z_A(V,T)$

$$Z_A(V,T) = \sum_{\vec{n}} \prod_k \frac{(x_k)^{n_k}}{n_k!} \tag{2}$$

The partition vector $\vec{n} = \{n_1, n_2, n_3, ....n_A\}$ is a partition of $A$ which is subject to the constraint $A = \sum k n_k$. Another equivalent representation of the vector $\vec{n}$ is $1^{n_1} 2^{n_2} 3^{n_3}...A^{n_A}$.

The vector $\vec{n}$ represent the division of an object of size $A$ into groups of varying sizes with $n_k$ of size $k$. Fig.1 shows the various partitions of $A = 1-7$ built on a lattice. Each partition can be represented by a box picture also shown in the figure.

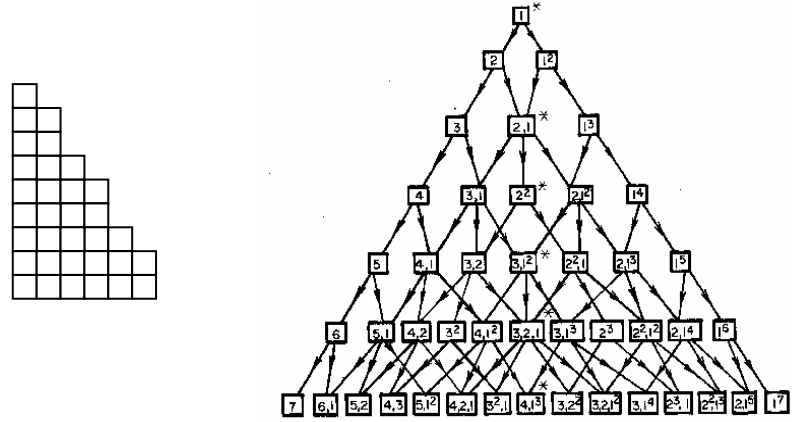

FIG.1. A block partition and a lattice of partitions. A block picture of a partition is shown in the left part of Fig.1. The vertical number of boxes or blocks in a given column is $k$ and the number of columns with k boxes or blocks is $n_k$. If an additional block is added to a given block picture in all possible positions, new partitions evolve with the connections shown in the right part of the figure. In the right part of the figure the $A$ ranges from 1 to 7 in the vertical direction. The partitions of a given $A$ appear in a given row. Each row has one unit higher $A$ than the previous row. The arrows indicate how each row is connected to the previous row by adding a single block.

The canonical partition function can be solved by a recurrence relation[22] given by

$$Z_A = \frac{1}{A}\sum_{k=1}^{A} kx_k Z_{A-k} \qquad (3)$$

and starting with $Z_0 = 1$. The grand canonical partition function $\mathcal{Z}(z,T)$ is

$$\mathcal{Z}(z,T) = \sum_{A=0}^{\infty} z^A Z_A(V,T) = \exp\left(\sum_{k=1}^{\infty} z^k x_k\right) \qquad (4)$$

where $z$ is the fugacity which is related to the chemical potential $\mu$ by $z = \exp\mu/k_B T$. The equation of state (EOS) is

$$\frac{PV}{k_B T} = \frac{\partial}{\partial V}\ln \mathcal{Z}(z,T) = \sum_{k=1}^{\infty} z^k x_k \qquad (5)$$

while the mean $A$ in the grand canonical ensemble is

$$A = z\frac{\partial}{\partial z} \ln Z(z,T) = \sum_{k=1}^{\infty} z^k kx_k \tag{6}$$

The fugacity of a system of fermions is determined by the following equation:

$$\frac{A}{V} = \rho \frac{2S+1}{\lambda_T^3}(z - \frac{z^2}{2^{3/2}} + \frac{z^3}{3^{3/2}} - ....) \tag{7}$$

The above equation can be inverted to give $z$ as a power series in $A$. The resulting virial expansion of the pressure is a power series expansion in $A$ which also involves the $x_k$'s as coefficients. The EOS virial expansion coefficients have an increasingly complex structure in terms of the $x_k$'s with increasing $A$. Specifically,

$$\frac{PV}{k_B T} = A - \frac{x_2}{x_1^2}A^2 + \frac{4x_2^2 - 2x_1 x_3}{x_1^4}A^3 + \frac{-20x_2^3 + 18x_1 x_2 x_3 - 3x_1^2 x_4}{x_1^6}A^4$$
$$+ \frac{-4x_1^3 x_5 + 112x_2^4 + 18x_1^2 x_3^2 - 144x_1 x_2^2 x_3 + 32x_1^2 x_2 x_4}{x_1^8}A^5 + .... \tag{8}$$

Substituting the results of Eq.(1) gives the following equation for fermions:

$$\frac{P}{k_B T} = \rho(1 + .1766\frac{\rho\lambda_T^3}{2S+1} - 3.3\cdot 10^{-3}(\frac{\rho\lambda_T^3}{2S+1})^2 + 1.11\cdot 10^{-4}(\frac{\rho\lambda_T^3}{2S+1})^3 +$$
$$- 3.5\cdot 10^{-6}(\frac{\rho\lambda_T^3}{2S+1})^4 + ...) \tag{9}$$

A rapid decrease in the numerical coefficients of ~1/30 allows an expansion around the ideal gas limit for relatively low temperatures compared to the Fermi temperature of $k_B T = E_F \sim 35 MeV$. For a system of fermions the first correction to the ideal gas value from pair correlation raises the pressure and associated energy due to the Pauli exclusion principle. The numerical coefficients come from $x_k = (-1)^{k+1}/k^{5/2}, k = 1,2,3,4$ with the value of $k$ representing the length of the cycle correlation coming from anti-symmetrizing the wave function. A factor such as $1/(2^{5/2})^2$ for $k = 2$ represents two different pair correlations with each pair correlation being a cycle of length two. The terms that contribute to a given power of $A$, say $A^n$ are the partitions of multiplicity $m = n-1$ in the partition $2(n-1)$. For example, for $n = 3$ the value of $2(n-1) = 4$. The partitions of 4 are 1+1+1+1, 2+1+1, 2+2, 3+1, 4. The partitions of 4 with multiplicity $m = n-1 = 2$ are then 2+2 and 3+1. Thus the numerator in front of $A^3$ has terms involving $x_2^2$ and $x_1 x_3$. The number of terms in front of $A^n$ increases exponentially fast. For $n = 16$, the number of terms of the partition of 30 into 15 parts is 176 which equals the number of integer partitions of 15. In general, the number of terms that appear in front of the coefficient of $A^{n+1}$ is equal to the number of partitions $P[n]$ of $n$. Asymptotically, the number of partitions of $n$ goes as

$$P[n] = \frac{e^{\pi\sqrt{2n/3}}}{4\pi\sqrt{3n}} \tag{10}$$

which is a Hardy-Ramanujan result. It is important to note that large cancellations can occur in higher order virial terms. The above example shows this cancellation is very large for symmetrization and antisymmetrization effects.

The index $k$ in a cluster representation of the partition function refers to the size a cluster. The pressure for a given $A^n$ is determined by the multiplicity and is independent of the mass of the cluster. While heavier cluster have a greater momentum change upon striking a wall, they move corresponding slower. Equipartition of energy gives $m<v^2>/2 = 3k_B T/2$ so that the pressure per particle is independent of the mass of any particle. Some further observations regarding specific terms that appear in front of $A^n$ are the following. In a virial expansion, the dimmer or cluster of size 2, appears in every coefficient (except the first term) as an term which is $x_2^{n-1}$. At the other extremes in size index $k$ the largest $k$ appears as $-(n-1)x_1^{n-2}x_n$. The $x_1^{n-2}$ brings the multiplicity to a total of $(n-2)+1 = n-1$. The next largest $k$ is $k = n-1$. The remaining factors must have a total multiplicity of $n-2$ and must involve $k's$ that sum to $2(n-1)-(n-1) = n-1$. Thus this term involves $x_1^{n-3}x_2 x_{n-1}$. A large k in the range $n \geq k \geq n/2+1$ for $n$ even and $n \geq k \geq (n+1)/2$ for $n$ odd can appear at most once because of the partition constraint. As a final remark for a reader familiar with Huang's statistical mechanics textbook [31], the $x_k$ is related to the $b_k$ in Eq.(10.27) of Ref [31] through the simple connection $x_k/x_1 = b_k$. The $b_k$ are the cluster integrals of the classical partition function of chapter10 which has a linked cluster diagrammatic expansion. The $b_k$ also appear in quantum cluster terms in the quantum case.

Fig. 2 shows various situations that appear in the evaluation of a partition function. The second virial coefficient, the term in front of $A^2$ in the equation of state, is determined by cycles of length two amongst monomers and bound clusters of size or continuum correlations of pairs of particles.

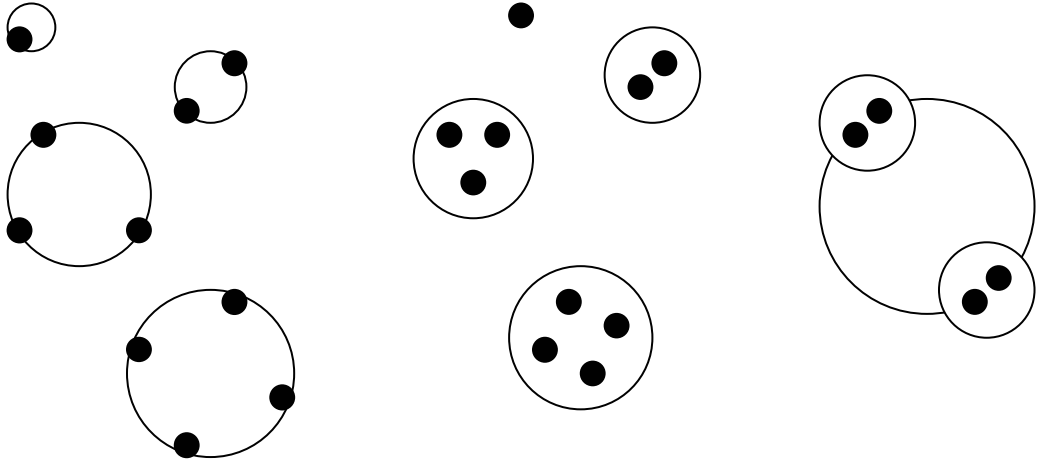

FIG. 2. Cycles and clusters that appear in the partition function. The left part of Fig. 2 shows a cycle class picture which arises from antisymmetrization effect for fermions or symmetrization effects for bosons. Unit cycles are represented by a single black dot with a circular loop, cycles of length two are two black dots connected by a ring. Cycles of length three and four are also shown. The middle part of the figure is a cluster view, with the size of the cluster determined by the number of black dots grouped inside the outer circle. The right part of this figure shows a cycle of length two that arises for exchange of two clusters of size two. Besides bound clusters, continuum correlations are also present.

The above results pertaining to one component systems can be generalized to two or more components [32]. For two components as in the nuclear case, the $x_k$'s are replaced with $x_{jk}$'s were $j$ refers to one of the components and $k$ the other component. If the index $j$ refer to the protons with total number $Z_p$ and the index $k$ for neutrons with total number $N_n$, the canonical partition function has a structure

$$Z_{Z_p,N_n}(V,T) = \sum_{\vec{n}_{jk}} \prod_k \frac{(x_{jk})^{n_{jk}}}{n_{jk}!} \tag{11}$$

The two constraints on the two component partition vector $\vec{n}_{jk}$ are then

$$Z_p = \sum_{j,k} j n_{jk}, \quad N_n = \sum_{j,k} k n_{jk} \tag{12}$$

These two constraints can also be summed to given a total baryon number constraint with $A = Z_p + N_n$, or can be subtracted to included a isospin constraint $2I_3 = (N_n - Z_p)/2$. The canonical partition function can be solved by any one of four equivalent recurrence relations. One relation based on proton number constraint reads

$$Z_{Z_p,N_n}(V,T) = \frac{1}{Z_p} \sum_{j,k} j x_{jk} Z_{Z_p-j,N_n-k} \tag{13}$$

*The second virial coefficient for an interacting gas with two components*

The thermodynamic behavior of the interacting gas with protons and neutrons can be developed by considering the following model. The EOS in an interacting gas to second order in the density is

$$\frac{PV}{k_B T} = A - b_2 A^2 \tag{14}$$

where the coefficient $b_2 \sim 1/V$ (not to be confused with the $b_k$ in Ref. [31] mentioned above) has components from the $nn, pp$ and $np$ terms. To see the structure of the second virial coefficient $b_2$ a simple example will be given where all three systems can form a single $S$-wave bound state, with the $np$ system having two spin possibilities. This simple idealized model can then be corrected for continuum interactions and the $nn$ and $pp$ bound states can be turned off to characterize the real system. The law of partial pressures leads to an EOS that is

$$\frac{PV}{k_B T} = N_p + N_n + N_{nn} + N_{pp} + N_{np} \tag{15}$$

The $Z = N_p + N_{np} + 2 N_{pp}$ and $N = N_n + N_{np} + 2 N_{nn}$ and $A = Z + N$. The number of $(ij)$ pairs ($i = p,n$ & $j = p,n$) is given by the Saha equation or law of mass action[18]

$$N_{ij}(S) = \frac{2S+1}{2^2} 2^{3/2} \frac{\lambda_T^3}{V} \exp[\frac{E_B(ij,S)}{k_B T}] N_i N_j \equiv a_2(ij,S) N_i N_j \tag{16}$$

The $\lambda_T = h/\sqrt{2\pi m k_B T}$ is the thermal wave length for a nucleon of mass $m$. The $2^{3/2}$ factor arises from the mass of a pair $m_{ij} = 2m$ and comes from the quantum wavelength $\lambda_T(ij) = h/\sqrt{2\pi m_{ij} k_B T}$. The $S$ is the spin of the $(ij)$ pairs and $E_B(ij)$ equals the binding energy of the pair. The Saha equation relates the number of $N_{ij}$ pairs to the existing equilibrium numbers of $N_i$ and $N_j$ and not the primordial numbers $Z, N$. Writing $N_p = Z - N_{np} - 2N_{pp}$ and $N_n = N - N_{np} - 2N_{nn}$ and defining the proton fraction $y_p$ as $y_p = y = Z/A$ and neutron fraction as $y_n = N/A = 1 - y_p = 1 - y$

leads to an EOS to order $A^2$ that is

$$\frac{PV}{k_B T} = A - \sum_S \sum_{i \& j = p,n} y_i y_j a_2(ij, S) \, A^2 \tag{17}$$

Exchange or antisymmetrization corrections in the $pp$ and $nn$ channels gives rise to further terms to be added to the EOS resulting in

$$\frac{PV}{k_B T} = A + \frac{1}{2^{5/2}} \frac{\lambda_T^3}{2V} (y^2 + (1-y)^2) A^2 - \sum_S \sum_{i \& j = p,n} y_i y_j a_2(ij, S) A^2 \tag{18}$$

As noted the $pp$ and $nn$ channels have no bound states, nor does the $np$ $S = 0$ channel. However a very long lived narrow resonance acts as a bound state and makes a contribution to the $A^2$ term through a term due to Beth and Ulhenbeck[33,18] that changes the bound state Boltzmann factor to a continuum correlation factor via a term involving the rate of change of the phase shift. Specifically, for $S$ – waves:

$$Exp[\frac{E_B}{k_B T}] \rightarrow \frac{1}{\pi} \int \frac{d\delta_0}{dk} \exp(-\hbar^2 k^2 / 2\mu k_B T) dk \tag{19}$$

where $\delta_0$ is the $S$ – wave phase shift and $\mu = m/2$ is the reduced mass. Higher orbital angular momentum correlations can also contribute through additional terms where $d\delta_0 / dk \rightarrow \Sigma_l (2l+1) \, d\delta_l / dk$ in the above equation. The integral over $k$ can also be rewritten as

$$\frac{1}{\pi} \int \frac{d\delta}{dE} \exp(-E/k_B T) dE = \frac{1}{\pi} \int \frac{\delta(E)}{k_B T} \exp(-E/k_B T) dE \tag{20}$$

The integral over phase shifts has the following special interesting cases:
i) If the interaction produces a resonance with energy $E_R$ in a channel with angular momentum $J_R$ which is narrow compared to the thermal energy $k_B T$, then the internal partition function has a contribution

$$(2J_R + 1) \exp(-E_R / k_B T) \tag{21}$$

This follows from the fact that a sharp resonance produces a rapid increase in a phase shift by $\pi$.
ii) A decreasing phase shift, ($d\delta_l / dk < 0$), will make a negative contribution to the continuum integral.
iii) An echo of a resonance, for which the phase shift decreases by $\pi$, results in a negative

contribution to the internal partition function. Moreover from Levinson's theorem [34], the difference in phase shift is given by $\delta_l(0) - \delta_l(\infty) = N_l \pi$, where $N_l$ equals the number of bound states with angular momentum $l$. At very high temperatures, echoes of resonances cancel the bound states and resonance contributions so that the system again behaves as an ideal gas. However, at moderate temperatures, a complete cancellation does not occur owing to the difference in energies of the bound states, resonances, echoes of resonances and from the fact that the rate of descent of the phase shift is limited by Wigner's limit which is $d\delta_l/dk > -R$ [18]. It should be noted that Levinson's theorem does not apply for potentials with an infinite hard core.

*Interaction energy*

For an ideal gas the internal energy is independent of the volume. The volume dependence of the energy can be obtained from the thermodynamic relation

$$\left.\frac{\partial E}{\partial V}\right|_T = T \left.\frac{\partial P}{\partial T}\right|_V - P \tag{22}$$

For an EOS of the form $P = k_B T (A/V - \hat{b}_2 A^2/V^2)$ with $\hat{b}_2/V = b_2$ the volume dependence of the energy is

$$E(V) \equiv E(\hat{b}_2) = (T \frac{\partial \hat{b}_2}{\partial T}) k_B T \frac{A^2}{V} \tag{23}$$

The second virial coefficient $\hat{b}_2$ is

$$\hat{b}_2 = \hat{b}_{2,exc} + \hat{b}_{2,int} = -\frac{\lambda_T^3}{2^{7/2}}(y^2 + (1-y)^2) + \sum_S \sum_{i \& j = p,n} y_i y_j \frac{2S+1}{2^2} 2^{3/2} \lambda_T^3 [\sum_B \exp(\frac{E_B(ij,S)}{k_B T}) +$$

$$\frac{1}{\pi} \sum_l (2l+1) \int \frac{\partial \delta_l(ij,S)}{\partial k} \exp(-\hbar^2 k^2/mk_B T) dk )] \tag{24}$$

The $\hat{b}_{2,exc}$ is the exchange part of $b_2$ which is the first term on the right hand side of this equation. The $1/2^{7/2} = 1/(2^{5/2}2)$ includes the spin degeneracy factor of 2. The $\hat{b}_{2,int}$ is the interaction part of $\hat{b}_2$ or the second term coming from binding energy terms and the third term is the continuum contribution.

The volume dependence in $E(V)$ comes from antisymmetrization terms for fermions (symmetrization terms for bosons) and from interaction terms with the latter called the interaction energy. The interaction energy density has both bound and continuum contributions and is

$$\varepsilon_{int} = \frac{E(\hat{b}_{2,int})}{V} = \frac{3}{2}k_B T \frac{A^2}{V^2} \lambda_T^3 \sum_S \sum_{i\&j=p,n} y_i y_j \frac{2S+1}{2^2} 2^{3/2} \left( -B_{b,c}(ij,S) + \frac{2}{3}T \frac{\partial B_{b,c}(ij,S)}{\partial T} \right) \quad (25)$$

where $B_{b,c}(ij,S) = B_b(ij,S) + B_c(ij,S)$ with the bound contribution given by

$$B_b(ij,S) = \sum_B (2J+1) \exp(\frac{E_B(ij,S)}{k_B T}) \quad (26)$$

and the continuum contribution given by

$$B_c(ij,S) = \frac{1}{\pi} \sum_l (2l+1) \int \frac{\partial \delta_l(ij,S)}{\partial k} \exp(-\hbar^2 k^2 / m k_B T) dk \quad (27)$$

A rescaled interaction energy density is defined as

$$\hat{\varepsilon}_{int} = \frac{\varepsilon_{int}}{\frac{3}{2}k_B T \frac{A^2}{V^2} \lambda_T^3 \frac{2^{3/2}}{4}} \quad (28)$$

Some features of the interaction density were studied in Ref [27].
The EOS is not only important in understanding features associated with the interaction energy but also the isothermal compressibility and entropy. The next subsection illustrates these features.

*Interaction effects and isothermal compressibility and entropy*
A quantity also related to the equation of state is the isothermal compressibility

$$\kappa_T = -\frac{1}{V}\frac{\partial V}{\partial P}\bigg|_T = \frac{1}{\frac{A}{V}k_B T - 2\hat{b}_2 \frac{A^2}{V^2} k_B T} = \frac{1}{P - \hat{b}_2 \frac{A^2}{V^2} k_B T} \quad (29)$$

Thus the sign of $\hat{b}_2$ determines whether $\kappa_T$ is above or below the ideal gas limit. If $\hat{b}_2$ is positive, interaction effects are more important than fermionic antisymmetrization effects and $\kappa_T$ will have a peak as the temperature is increased from low to high temperatures as discussed in Ref [35]. Higher order interaction effects coming from higher order clusters and the presence of a hard core potential are also important in understanding the compressibility of nuclear matter. Repulsive components are necessary for saturation of cold nuclear matter at the proper density. Calculations of nuclear mater on a lattice show a peak in the compressibility [16].

Properties of the entropy can also be related to the equation of state. For example the

Helmholtz free energy $F = U - TS$, ($U$ is equal to $E$) with $dF = -SdT - PdV + \mu dA$ leads to a Maxwell relation: $(\partial S/\partial V)_T = -(\partial P/\partial T)_V|$. The second law $TdS = dQ$ and the heat capacity at constant volume $C_V$, with $dQ = C_V dT = dU$ can be used to find the temperature dependence of $S$ at constant $V$. The Euler equation $U = ST - PV + \mu A$ or $S = (U + PV + \mu A)/T$ and $dU = TdS - PdV + \mu dA$ leads to the Gibbs-Duheim equation $SdT - VdP + Ad\mu = 0$ and $S = (\partial P/dT)_\mu$. For a monatomic ideal gas of $A$ nucleons, the entropy is given by the Sakur-Tetrode expression [18]:

$$S = S_{id} = Ak_B \ln e^{5/2} \frac{Vg_S}{A\lambda_T^3} \tag{30}$$

Antisymmetrization effect for fermions and symmetrization effects for bosons lead to correction to the Sakur-Tetrode law that reads

$$S = Ak_B \ln e^{5/2} \frac{Vg_S}{A\lambda_T^3} \pm \frac{1}{2^{7/2}} \frac{A\lambda_T^3}{g_S V} + 2(2^{-4} - 3^{-5/2})(\frac{A\lambda_T^3}{g_S V})^2 + \ldots \tag{31}$$

with the + sign for fermions and the –sign for bosons in the ± term. Including interaction terms up to the second order virial coefficient, where $PV = Ak_B T(1 - \hat{b}_2(A/V))$, leads to a correction to $S$ which is

$$S - S_{id} \equiv S(\hat{b}_2) = \frac{A^2}{V} k_B \frac{d(T\hat{b}_2)}{dT} = \frac{A^2}{V} k_B \{T \frac{d\hat{b}_2}{dT} + \hat{b}_2\} \tag{32}$$

The first term in curly brackets appears in the energy of Eq.23 while the second term in these brackets is the correction to the pressure from $\hat{b}_2 = Vb_2$ (Eq.14 with a positive sign). The chemical potential associated with $\hat{b}_2$ can be obtained from $U = ST - PV + \mu A$ or from the Helmholtz free energy $F = U - TS$ using $\mu = \partial F/\partial A$ with $V, T$ constant. The chemical potential $\mu(\hat{b}_2) = -2k_B T\hat{b}_2 A/V$ and the free energy associated with $\hat{b}_2$ is given by $F(\hat{b}_2) = -k_B T\hat{b}_2(A^2/V)$. The $\hat{b}_2$ corrections to the isothermal compressibility in the unitary limit in Eq.(29) can be rewritten as

$$\kappa_T = \frac{\kappa_{T.ideal}}{1 - 2\hat{b}_2 \frac{A}{V}} = \frac{\kappa_{T.ideal}}{1 - \frac{\mu(\hat{b}_2)}{k_B T}} \approx \kappa_{T.ideal} \left(1 + \frac{\mu(\hat{b}_2)}{k_B T}\right) \tag{33}$$

The $\kappa_{T.ideal} = 1/((A/V)k_B T)$ is the ideal gas compressibility. Once $\hat{b}_2$ is given, the $S(\hat{b}_2)$, $\mu(\hat{b}_2)$, $F(\hat{b}_2)$, $E(\hat{b}_2)$ and $\kappa_T$ can all be calculated.

II.B *Interaction Potentials*

To proceed in a study of interaction effects, an interaction potential is introduced. The nuclear force has a short range repulsive part besides an attractive longer range part. As an approximation to a real system with these features, a simplified interaction which an infinite repulsive core in the range $0 \leq r \leq c$ with an attractive square well of depth $V_0$ in the range $c \leq r \leq R$ will be used in this study. Moreover, various limits can be made using this potential. These limits include the following.

1. The hard core Bose gas and hard core Fermi gas where the attractive square well part is turned off. Both the hard sphere Bose gas and the hard sphere Fermi gas appear in textbooks and a comparison can thus be made with these examples as a limiting situation. These results are given in Appendix A
2. A pure attractive square well is a limit with no hard core. The Ho/Muller[6,7] study in atomic systems uses only an attractive potential and their results are for a one component system.
3. A delta shell potential which has no hard core and has all the interaction occurring on a shell where $c \to R \equiv a$ and the $V_0 \to \infty$. This potential was used by Gottfried as an example for scattering theory in quantum mechanics [36]. Results for the delta shell potential $V(r) = -\chi \delta(r-a)$ are developed in Appendix B and comparisons are made there.

*Attractive square well plus repulsive hard core potential*
*S – wave phase shift*
The $S$ – phase shift for a square well potential with a repulsive core is

$$\tan(\delta_0) = \frac{k \tan \alpha (R-c) - \alpha \tan kR}{\alpha + k \tan kR \cdot \tan \alpha (R-c)} \tag{34}$$

or: $\delta_0 = \arctan[(kR/\alpha R)\tan \alpha(R-c) - k(R-c) - kc$ and

$$\frac{\partial \delta_0}{\partial k} = \frac{\alpha_0^2 (\alpha \tan \alpha(R-c) - \alpha(R-c))}{\alpha[(\alpha^2 + (k \tan \alpha(R-c))^2]} - c \tag{35}$$

The $\alpha^2 = k^2 + \alpha_0^2$ and $\alpha_0 = \sqrt{2\mu |V_0|/\hbar^2}$. Fig. 3 shows the behavior or $\partial \delta_0 / \partial k$ with $k$.

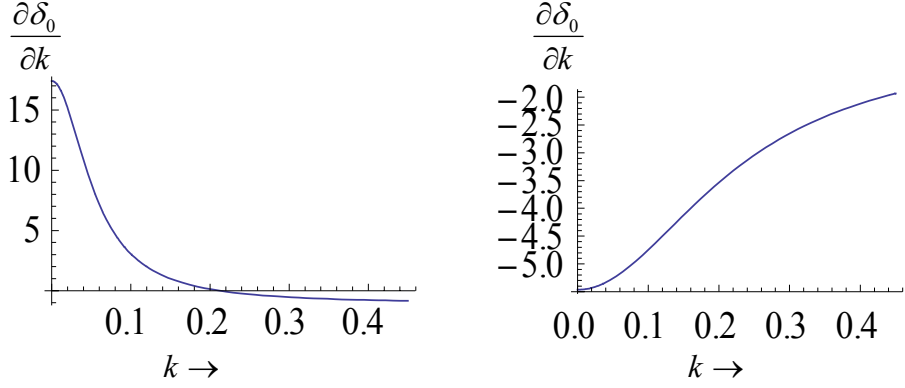

FIG. 3. Behavior or $\partial \delta_0 / \partial k$ with $k$. The left part of the figure is for the case when the potential has no bound state. The $V_0 = 31.6 MeV, R = 2 fm, c = 0.27 fm$ were used to generate the behavior shown. The right figure is for the case when the potential has a bound state energy $E_B = 2.2 MeV$. The $V_0 = 57.14 MeV, R = 1.8 fm, c = 0.27 fm$. These choices of $V_0, R, c$ fit low energy scattering data for $nn$ scattering data(left figure) and $np$ spin $S = 1$ scattering data as will be discussed below.

The $\sin^2 \delta_0$ behavior is shown in Fig.4.

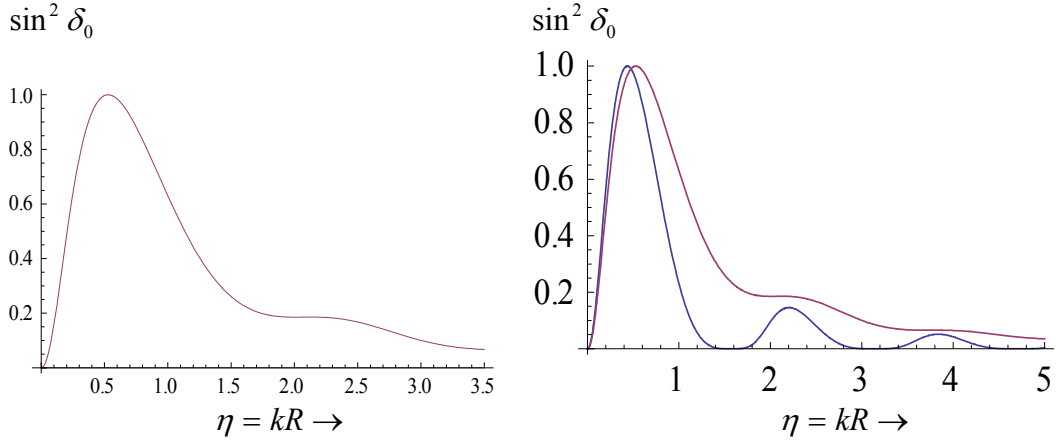

FIG. 4. The behavior of $\sin^2 \delta_0$ with $kR$. The left figure is the square well potential result. The $\alpha_0 = 1/fm, R = 2 fm$ were used to generate the curve. Since $\alpha_0 R = 2 > \pi/2$ the square well has a $S-$wave bound state. The first large bump is not a true resonance. The right figure also contains the delta-shell result adjusted to have the same bound state energy and radius of the square well used in the left figure.

An attractive Dirac delta function potential is obtained as a limiting situation with $|V_0| \to \infty, R \to 0, |V_0| R^2 \to K_\delta$. Such a potential can have at most one bound state which is in a $S-$wave channel. A surface delta or delta shell interaction, whose properties are discussed in appendix A, also has only one $S-$wave bound state but can have bound states in higher angular momentum states depending on its strength and radius of the

interaction. The surface delta potential is $V(r) = -\chi\delta(r-a)$..

*Effective range theory limit*

When $k \ll \alpha_0$ the behavior of the phase shift is given by an effective range theory. Specifically, for $S$-waves:

$$k \cot \delta_0 \cong -\frac{1}{a_{sl}} + \frac{1}{2}r_0 k^2 \tag{36}$$

with $a_{sl}$ the scattering length and $r_0$ the effective range. In general for a partial wave with angular momentum $l$ the effective range approximation is $k^{2l+1} \cot \delta_l \cong -1/a_{sl}^l + r_0^l k^2/2$. For $l=0$, the upper subscript referring to $l$ is omitted. For a square well with hard core the $a_{sl}$ and $r_0$ are given by

$$\begin{aligned} a_{sl} &= R\left(1 - \frac{\tan\alpha_0(R-c)}{\alpha_0(R-c)}\right) \\ r_0 &= R - \frac{1}{\alpha_0^2 a_{sl}} - \frac{1}{3}\frac{R^3}{a_{sl}^2} + c\left(1 - \frac{2R}{a_{sl}} + \frac{R^2}{a_{sl}^2} + \frac{1}{\alpha_0^2 a_{sl}^2}\right) \end{aligned} \tag{37}$$

In the effective range approximation the rate of change of the phase shift is

$$\frac{d\delta_0}{dk} = -\frac{1}{1+\cot^2\delta_0}\left(\frac{1}{a_{sl}k^2} + \frac{r_0}{2}\right) \tag{38}$$

Substituting the effective range approximation into this last equation gives the exact effective range result that is:

$$\frac{d\delta_0}{dk} = -\frac{a_{sl}}{1 + a_{sl}(a_{sl}-r_0)k^2 + \frac{1}{4}(r_0 a_{sl})^2 k^4}\left(1 + \frac{r_0 a_{sl}}{2}k^2\right) \tag{39}$$

In the limit that the scattering length $|a_{sl}| \gg r_0$, the effective range, and when $k$ is small so that $r_0 a_{sl} k^2 \ll 1$ the following simplified result is obtained:

$$\frac{d\delta_0}{dk} \approx -\frac{a_{sl}}{1 + a_{sl}(a_{sl}-r_0)k^2}\left(1 + \frac{r_0 a_{sl}}{2}k^2\right) \tag{40}$$

In this limit, the integral over $d\delta_0/dk$ can be done analytically and is of a form

$$B_c \equiv B_c(S-\text{wave}) = \frac{1}{\pi}\int_0^\infty \frac{d\delta_0}{dk}\exp(-bk^2)dk =$$

$$-\frac{a_{sl}^2(2a_{sl}-3r_0)}{4(a_{sl}^2-a_{sl}r_0)^{3/2}}\exp(\frac{b}{a_{sl}^2-a_{sl}r_0})Erfc(\sqrt{\frac{b}{(a_{sl}^2-a_{sl}r_0)}}) - \frac{a_{sl}^2 r_0}{4(a_{sl}^2-a_{sl}r_0)\sqrt{\pi b}} \quad (41)$$

The Boltzmann exponential factor $\exp(-bk^2)$ suppresses the $k^4$ term in $\partial\delta_0/\partial k$ and makes Eq. (41) and very good approximation to the complete effective range result for it. The $b = \hbar^2/2\mu k_B T = \lambda_T^2(\mu)/2\pi$ and $Erfc(\sqrt{b/a_{sl}^2}) = 1 - Erf(\sqrt{b/a_{sl}^2})$. In the limit that $a_{sl} \gg r_0$ and with $(r_0 a_{sl})^2 k^4$ terms included, the $B_c$ is simply

$$B_c = -sign[a_{sl}]\frac{1}{2}\exp(\frac{b}{a_{sl}^2})Erfc(\sqrt{b/a_{sl}^2}) - \frac{1}{2}\exp(4\frac{b}{r_0^2})Erfc(2\sqrt{b/r_0^2}) \quad (42)$$

The $Erfc(x)$ has an expansion for small $x$ that reads $Erfc(x) = 1 - 2x/\sqrt{\pi}$. For large $x$ $Erfc(x) \sim \exp(-x^2)/(x\sqrt{\pi})$. For large $2\sqrt{b/r_0^2}$, the second term on the right hand side $-\exp(4b/r_0^2))Erfc(2\sqrt{b/r_0^2})/2 \to -r_0/4\sqrt{\pi b}$. The quantity $-B_c - (2/3)b(dB_c/db)$ appears in $\varepsilon_{\text{int}}$. Thus a $\Delta_E B_c$ is defined as $\Delta_E B_c \equiv B_c + (2/3)(bdB_c/db)$. When $a_{sl}$ becomes large then

$$-\Delta_E B_c \equiv \to \frac{1}{6}\frac{r_0}{\sqrt{\pi b}}(1+\frac{r_0}{a_{sl}}+(\frac{r_0}{a_{sl}})^2) + sign(a_{sl})(\frac{1}{2}-\frac{4}{3\sqrt{\pi}}\frac{\sqrt{b}}{a_{sl}}(1+\frac{r_0}{a_{sl}}) + \frac{5}{6}\frac{b}{a_{sl}^2} - \frac{3}{16}(\frac{r_0}{a_{sl}})^2)$$

$$\to (a_{sl} \to \infty) \quad sign(a_{sl})(\frac{1}{2}) + \frac{1}{6}\frac{r_0}{\sqrt{\pi b}} = sign(a_{sl})(\frac{1}{2}) + \frac{1}{6}\sqrt{2}\frac{r_0}{\lambda_T} \quad (43)$$

Thus the factor $r_0/\lambda_T$ appears as a correction to the universal thermodynamic limit of $sign[a_{sl}]/2$.

Fig. 5 makes a comparison of the behavior of the integral $B_c$ for the following situations. The values of $B_c$ are compared for the exact $S$-wave phase shift for a square well potential and the simple analytic expression for $B_c$ based on the approximate effective range approximation neglecting the $(r_0 a_{sl})^2 k^4$ term.

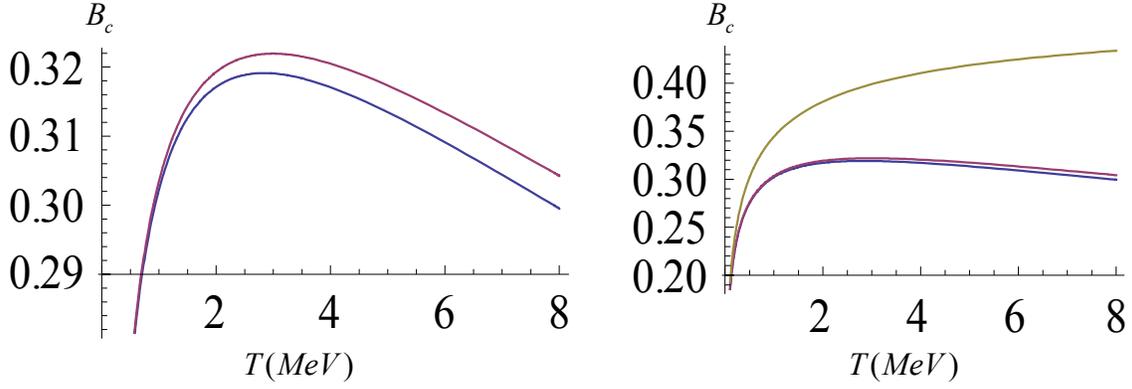

FIG. 5. $B_c$ versus temperature $T(MeV)$. Left figure. The upper curve is $B_c$ calculated using the approximate expression of Eq.(41) while the lower curve is the exact square well result. Very little error is present using the approximate simple analytic expression for $B_c$ over a wide range of temperatures. The value of the scattering length is $a_{sl}$ -17.4 $fm$ and the effective range $r_o = 2.4\, fm$. These values are appropriate for $nn$ scattering. The corresponding square well parameters are listed in Table1 and are $\{V_0, R, c\} = \{31.6\, MeV,\ R = 2.4\, fm, c = 0.27\, fm\}$. The scattering length $a_{sl} \rightarrow -\infty$ at $\{V_0, R, c\} = \{34.106\, MeV,\ R = 2.4\, fm, c = 0.27\, fm\}$. Right figure. The right figure now includes a calculation of $B_c$ (the upper curve) with the effective range $r_o$ set equal to zero and has $a_{sl}$ -17.4 $fm$. The two lower curves are the same as in the left figure. The importance of including a scattering length is seen in the comparison.

Fig. 6 shows that by rescaling the effective range approximation of $B_c$ calculated from Eq.(41) that a resulting rescaled curve very accurately follows the exact square well result.

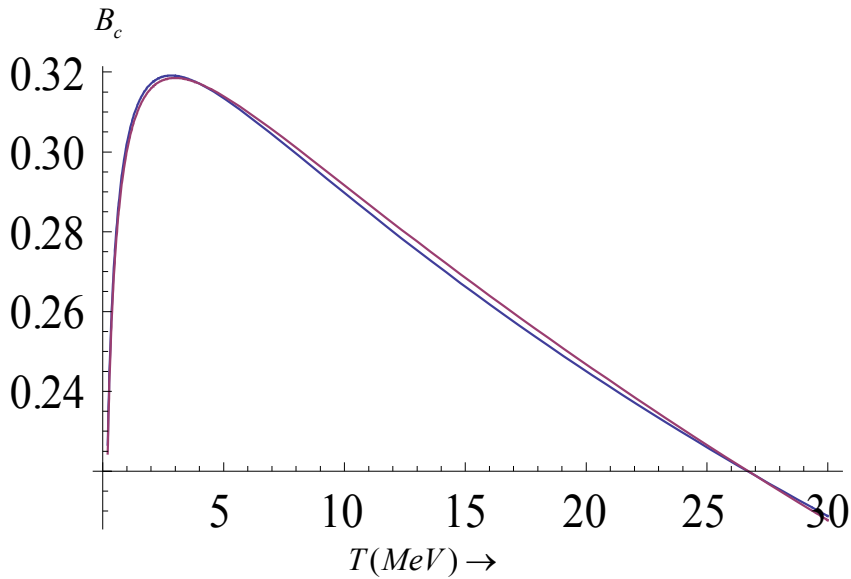

FIG. 6. The quantity $B_c$ calculated using a square well potential and a rescaled effective range potential result. The rescaling factor is chosen so as to have both values agree at $T = 4 MeV$. The rescaling factor is .9893 at this temperature. The slightly upper curve at high temperatures ~10-20 $MeV$ is the simple effective range result of Eq.(41). The ratio of the two values at $T = 20 MeV$ is only 1.007 showing excellent agreement even at very high temperatures between the simple effective range result and the exact square well result.

The quantity $|\Delta_E B_c|$ versus temperature is shown in Fig.7. The universal thermodynamic limit for $\Delta_E B_c$ is $sign(a_{sl})(1/2)$. Effective range corrections to $\Delta_E B_c$ are given in Eq.(43).

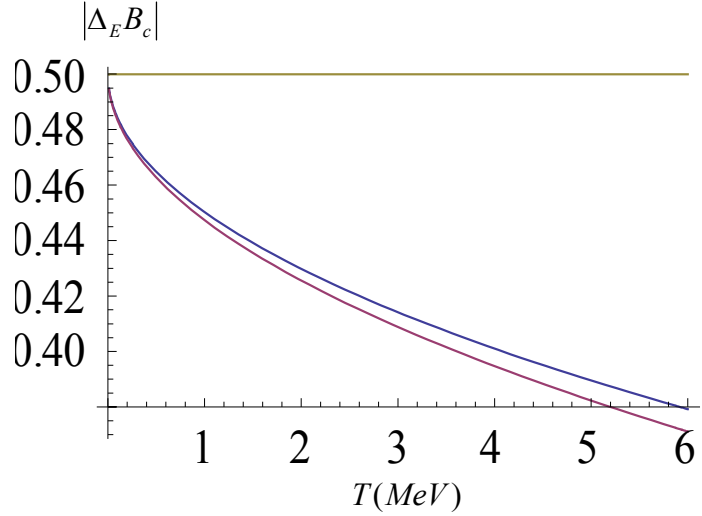

FIG 7. The quantity $|\Delta_E B_c|$ versus temperature. The figure shows the importance of including effective range corrections in the unitary limit. The upper horizontal curve is the unitary limit neglecting effective range corrections. For the other two curves, the upper curve is $|\Delta_E B_c|$ calculated using the approximate expression of Eq. (43) while the lower curve is the exact square well result. Very little error is present using the approximate simple analytic expression of Eq.(43) for $B_c$ over a wide range of temperatures. The parameters are the same as in Fig. 6.

The $S-$wave effective range parameters are given in Table 1. The experimental parameters are from Ref [35]. Also shown are calculated values for a square well potential with a hard core repulsion.

TABLE 1. Experimental values of scattering lengths and effective ranges followed by square well paramenters and calculated values. The value $c = 0.27$ was used. The $np$ S=1 parameters bind the deuteron at 2.2 $MeV$. The $pp$ system has Coulomb terms with $a_{sl} = -7.821 fm$, $r_{0s} = 2.83\ fm$. The units of $a_{sl}, r_0, R$ given below are $fm$ with $V_0$ in $MeV$ and are omitted. More recent values of $a_{sl}$ & $r_0$ can be found in Ref [36]. The results given in this paper are insensitive to the precise values of these quantities which are similar in [35] and [36].

|  | np S=1 | np S=0 | nn S=0 |
|---|---|---|---|
| Exp: | $a_{sl} = a_t = 5.4$ | $a_{sl} = a_{s,np} = -23.7$ | $a_{sl} = a_{s,nn} = -17.4$ |
| Exp: | $r_0 = r_{0t} = 1.75$ | $r_0 = r_{0s,np} = 2.73$ | $r_0 = r_{0s,nn} = 2.4$ |
| $\{V_0, R\}$: | $\{57.14, 1.8\}$ | $\{23.18, 2.3\}$ | $\{31.60, 2.0\}$ |
| Cal: | $a_{sl} = a_t = 5.4$ | $a_{sl} = a_{s,np} = -23.70$ | $a_{sl} = a_{s,nn} = -17.4$ |
| Cal: | $r_0 = r_{0t} = 1.73$ | $r_0 = r_{0s,np} = 2.69$ | $r_0 = r_{0s,nn} = 2.40$ |

*Interaction entropy and the unitary limit*

The expression of Eq.32 and properties of $\hat{b}_2$ can be used to calculate the interaction entropy. The exchanged part of $\hat{b}_{2,exc} = -1/2^{7/2}(y^2 + (1-y)^2)$ from Eq.(24) gives

$$S(\hat{b}_{2,ech}) = \frac{1}{2^{9/2}} \frac{A\lambda_T^3}{V}(y^2 + (1-y)^2) \tag{44}$$

The interaction part of $\hat{b}_{2,int}$ of Eq.(24) can be used to calculate $S(\hat{b}_{2,int})$ using Eq.(32). As an example of an evaluation of $S(\hat{b}_{2,int})$ is a system of pure neutron with negative scattering length. Then only the continuum contributes and

$$\frac{S(\hat{b}_{2,int})}{\frac{A^2}{V} k_B \frac{2^{3/2}}{2^2}} = \frac{d}{dT}(T\lambda_T^3 B_C) = -\frac{1}{2}\lambda_T^3 B_C + \lambda_T^3 T \frac{dB_C}{dT} = \lambda_T^3\left(-\frac{1}{2}B_C - b\frac{dB_C}{db}\right) \tag{45}$$

Thus,

$$\frac{TS(\hat{b}_{2,int})}{\frac{A^2}{V} k_B T \frac{2^{3/2}}{2^2} \lambda_T^3} = \frac{1}{2}\left(-B_C - 2b\frac{dB_C}{db}\right) \tag{46}$$

which can be compared with the corresponding equation for the interaction energy
$E(\hat{b}_{2,int}) = V\varepsilon_{int}$

$$\frac{E(\hat{b}_{2,int})}{\frac{A^2}{V}k_B T \frac{2^{3/2}}{2^2}\lambda_T^3} = \frac{3}{2}\left(-B_C - \frac{2}{3}b\frac{dB_C}{db}\right) \qquad (47)$$

The abbreviated notation $B_C = B_C(nn, S=0)$ is used with $B_C$ given by Eq.42 as a very good approximation as discussed. Since $b = \lambda_T^2/2\pi \sim 1/T$ then $T(d/dT) = -b((d/db))$. The quantities $B_C + (2/3)b(dBc/dB) \equiv \Delta_E Bc$ and $B_C + 2b(dBc/dB) \equiv \Delta_S Bc$ are plotted in Fig.8 versus temperature in the unitary limit $a_{sl} \to -\infty$.

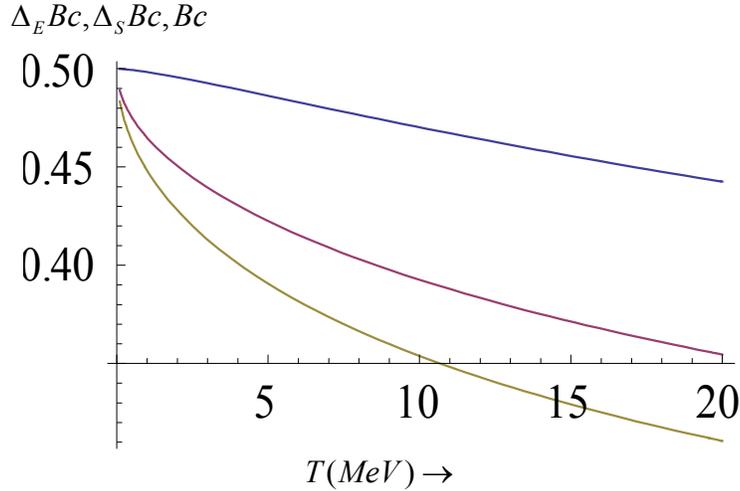

FIG.8. The behavior of $\Delta_E Bc, \Delta_S Bc$ and $Bc$ versus $T$. The calculation were done with $r_0 = 2.4\,fm$ and $a_{sl} \to -\infty$. The upper curve is $\Delta_S Bc \equiv B_C + 2b(dBc/dB)$ and the middle curve is $\Delta_E Bc \equiv B_C + (2/3)b(dBc/dB)$. The lowest curve is just $Bc$. The effective range corrections are much larger for the energy than the entropy. The effective range corrections for the entropy are rather small over a broad range of temperatures as shown in the figure. The energy also has a factor $3/2$ compared to the $1/2$ factor for the entropy on the right hand side of their respective equations. The Helmholtz free energy and chemical potential involve $Bc$ only and not a derivative of $Bc$.

As shown in Fig.8, $\Delta_S B_C$ has a much weaker dependence on $T$ than $\Delta_E B_C$. This feature arises from a cancellation between $B_C$ and $2b(dB_C/db)$. To understand this behavior, the large $x$ expansion of $\exp(x^2)Erfc(x)$ is needed to a higher order than $(1/x\sqrt{\pi})$. Namely,

$\exp(x^2) Erfc(x) = (1/x\sqrt{\pi})\{1 - 1/(2x^2)\}$ for large $x$ with $x = 2\sqrt{r_0^2/b}$ and $b = \lambda_T^2/2\pi$. The next higher order term inside the curly bracket $\sim (1/x^4)$ and is omitted. Then

$$\Delta_S B_C = \frac{1}{2} - \frac{1}{16\sqrt{\pi}}(\frac{r_0}{\sqrt{b}})^3 = \frac{1}{2} - \frac{\sqrt{2\pi}}{8}(\frac{r_0}{\lambda_T})^3 \tag{48}$$

and $\Delta_S B_C$ has no $r_0/\lambda_T$, $(r_0/\lambda_T)^2$, $(r_0/\lambda_T)^4$ and terms while

$$\Delta_E B_C = \frac{1}{2} - \frac{1}{6\sqrt{\pi}}\frac{r_0}{\sqrt{b}} - 0\cdot(\frac{r_0}{\sqrt{b}})^3 = \frac{1}{2} - \frac{\sqrt{2}}{6}\frac{r_0}{\lambda_T} \tag{49}$$

The $\Delta_E B_C$ has a $r_0/\lambda_T$ term but no $(r_0/\lambda_T)^2$, $(r_0/\lambda_T)^3$ and $(r_0/\lambda_T)^4$ terms. The interaction part of the Helmholtz free energy $F = E - TS$ is

$$\frac{F(\hat{b}_{2,int})}{\frac{A^2}{V}k_B T \frac{2^{3/2}}{2^2}\lambda_T^3} = -B_C \tag{50}$$

with

$$B_C = \frac{1}{2} - \frac{1}{4\sqrt{\pi}}\frac{r_0}{\sqrt{b}} + \frac{1}{32\sqrt{\pi}}(\frac{r_0}{\sqrt{b}})^3 = \frac{1}{2} - \frac{\sqrt{2}}{4}\frac{r_0}{\lambda_T} + \frac{3\sqrt{2\pi}}{16}(\frac{r_0}{\lambda_T})^3 \tag{51}$$

Effective range corrections to the ideal gas compressibility $\kappa_{T.ideal} = 1/((A/V)k_B T)$ in the unitary limit are given by Eq.(33). Since these corrections involve $B_C$ they contain both linear and cubic terms in the ratio $r_0/\lambda_T$ by Eq.(51).

*Square well - p - wave properties*
For a square well potential in the absence of a hard core the p-wave phase shift is

$$\tan\delta_1 = \frac{Rk\alpha^2 \tan\alpha R - Rk^2\alpha \tan kR - \alpha_0^2 \tan kR \tan\alpha R}{Rk^2\alpha + (\alpha_0^2)\tan\alpha R + Rk\alpha^2 \tan kR \tan\alpha R} \tag{52}$$

For a repulsive potential step, the $\alpha^2 = k^2 - 2\mu|V_0|/\hbar^2$. The hard core limit of a repulsive step is obtained by taking $|V_0| \to \infty$ and using $\tanh(z) = -i\tan(iz)$. The above equation for $\tan\delta_1$, with $\alpha_0^2$ replaced by $-\alpha_0^2$ for a repulsive step, can be rewritten as

$$\tan \delta_1 = \frac{\alpha^2 (Rk \tan \alpha R - Rk^2 (1/\alpha) \tan kR + (\alpha_0^2/\alpha^2) \tan kR \tan \alpha R)}{\alpha^2 (Rk^2(1/\alpha) - (\alpha_0^2/\alpha^2) \tan \alpha R + Rk \tan kR \tan \alpha R)} \to$$

$$\frac{\tan \alpha R (kR + (\alpha_0^2/\alpha^2) \tan kR)}{\tan \alpha R ((-\alpha_0^2/\alpha^2) + kR \tan kR)} \to \frac{(kR - \tan kR)}{1 + kR \tan kR} \tag{53}$$

The last step follows using $(\alpha_0^2/\alpha^2) \to -1$ in a pure repulsive hard core limit with core radius $R$. The presence of a hard core potential with radius $c$ modifies the expression for an attractive square well potential with radius $R$ and gives rise, for arbitrary $l$, to the result

$$\tan \delta_l = \frac{[xj'_l(x) j_l(y) - yj'_l(y) j_l(x)] - \lambda_l [xj'_l(x) \eta_l(y) - y\eta'_l(y) j_l(x)]}{[x\eta'_l(x) j_l(y) - yj'_l(y) \eta_l(x)] - \lambda_l [x\eta'_l(x) \eta_l(y) - y\eta'_l(y) \eta_l(x)]} \tag{54}$$

The $x = kR$, $y = \alpha R$ and $\lambda_l = j_l(\alpha c)/\eta_l(\alpha c)$. For $l=1$, $\lambda_1 = (\alpha c - \tan(\alpha c))/(1 + \alpha c \tan(\alpha c))$. As a check, the pure hard core limit is obtained by taking the attractive square well part of the potential in the limit $\alpha_0^2 \to 0, \alpha \to k, R \to c$ which leads to the result $\tan \delta_1 \to \lambda_1$. A attractive square well is simply obtained from the limit $\lambda_1 \to 0$.

Fig. 9 shows $P$-wave phase shifts for various types of potentials.

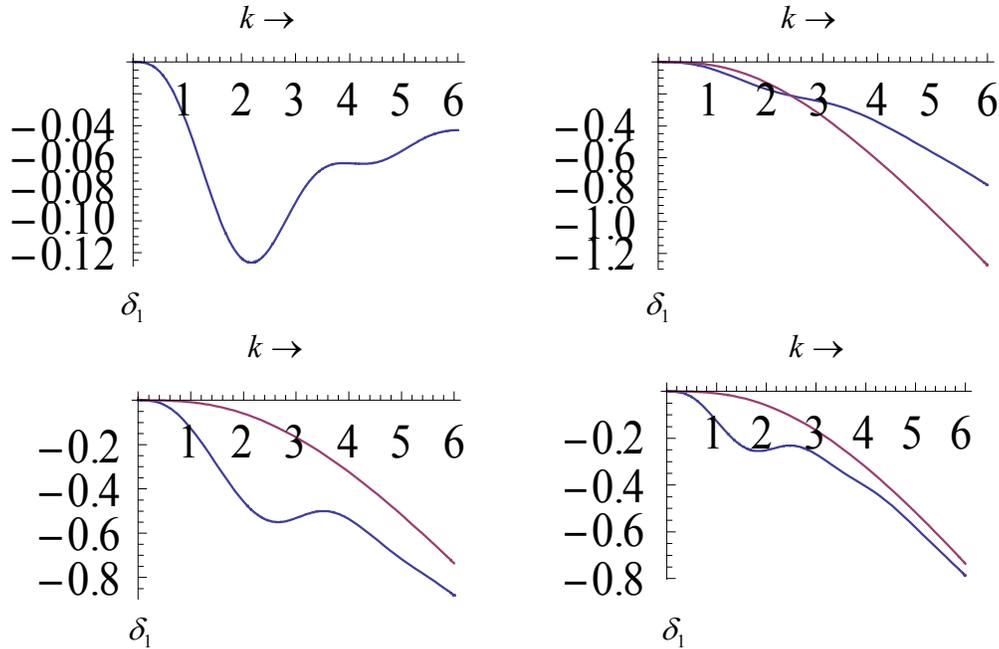

FIG. 9. Behavior of $P$-wave phase shifts versus $k$. The phase shift $\delta_1$ is in radians. The upper left part of the figure is for a purely repulsive square well potential with no hard core. The parameters used are $|\alpha_0| = .6/fm$, $R = 1.5 fm$. The phase shift $\delta_1$ goes to zero at high $k$ by Levinson's theorem since the well has no bound state or hard core. The upper

right part of the figure now also includes a hard core plus the repulsive step and pure hard core potential. A pure hard core potential with hard core radius $c = .3\,fm$ is represented by the lower curve at high $k$. The second curve is for a repulsive step with $|\alpha_0| = .6/\,fm$, $R = 1.5\,fm$ plus hard core with $c = .3\,fm$. The lower left part of the figure shows the effect of increasing the strength of the repulsive part of the potential to $|\alpha_0| = 1.2/\,fm$ keeping $R = 1.5\,fm$, $c = .3\,fm$. The phase shift has turned upward for a small domain in $k$ reflecting the behavior of in the upper left figure. Eventually, the hard core dominates in the higher region of $k$. The lower right curve shows the effect of increasing the radius and uses $|\alpha_0| = .6/\,fm$, $R = 2\,fm$, $c = .3\,fm$.

The presence of spin and isospin in a space anti-symmetric $P$-wave channel has phase shifts characterized as follows. The symmetric isospin $I = 1$ channel is associated with the symmetric spin $S = 1$ triplet and corresponding $^3P_0, ^3P_1, ^3P_2$ phase shifts with $J = 0, 1, 2$. The antisymmetric isospin $I = 0$ has the antisymmetric singlet $S = 0$ state and $^1P_1$ phase shift with $J = 1$. Both a nuclear spin-orbit force and tensor force are necessary to explain features associated with the $P$-wave phase shifts. Fig.10, taken from Bohr and Mottleson [39], shows various $S, P, D$-wave phase shifts. A spin-orbit force given by $V_{LS}(r)\vec{L}\cdot\vec{S} = V_{LS}(r)(J(J+1) - l(l+1) - s(s+1))/2$ cannot account for the behavior of the triplet $P$-wave phase shifts. If the spin-orbit were the only spin dependent force the $^3P_1, J = 1$ phase shift would be intermediate between the $^3P_2, J = 2$ and $^3P_0, J = 0$ phase shifts from the $J(J+1)$ dependence in $\vec{L}\cdot\vec{S}$.

The $J$-weighted average of the $P$-wave $S = 1$ phase shifts given by

$$\bar{\delta}(^3P) = \frac{1}{9}\{5\delta(^3P_2) + 3\delta(^3P_1) + 1\delta(^3P_0)\} \tag{55}$$

was noted to be small in Bohr and Mottleson. Table2 summarizes the behavior of the $P$-wave spin triplet $^3P_0, ^3P_1, ^3P_2$ and spin singlet $^1P_1$ phase shifts.

TABLE 2. Behavior of the $J$-weighted sum of the $P$-wave spin $S = 1$ and $S = 0$ phase shifts. The phase shifts are taken from Ref [40] of Arndt etal. Except for the case $^1P_1$, the error bars in each of the other phase shifts are small and therefore not given.

| Energy | State | $\delta$(deg) | $\bar{\delta}(^3P))$ | $3\delta(^1P_1), 9(\bar{\delta}(^3P))$ |
|---|---|---|---|---|
| 10. MeV | $^3P_0$ | 3.90 | | |
| | $^3P_1$ | -2.32 | | |
| | $^3P_2$ | 0.70 | | |
|         |           |              | 0.44/9~0.05  | .44          |
|         | $^1P_1$   | $-3.37\pm1.$ |              | $-10.11\pm3$ |
| 25. MeV | $^3P_0$   | 8.96         |              |              |
|         | $^3P_1$   | -5.31        |              |              |
|         | $^3P_2$   | 2.61         |              |              |
|         |           |              | 6.08/9~0.68  | 6.08         |
|         | $^1P_1$   | $-4.92\pm.43$|              | $-14.76\pm1.29$ |
| 50. MeV | $^3P_0$   | 12.74        |              |              |
|         | $^3P_1$   | -8.81        |              |              |
|         | $^3P_2$   | 6.02         |              |              |
|         |           |              | 16.41/9~1.82 | 16.41        |
|         | $^1P_1$   | $-4.23\pm.57$|              | $-12.69\pm1.71$ |

___

Thus a relatively weak central force in triplet $P$ states is present. A central Serber force of the type $V(r)(1+P^r)$, with $P^r$ a space exchange operator, would vanish in odd angular momentum states. Such a force also accounts for the large back angle scattering in the $np$ differential cross section.

The $^1P_1$ contribution is in the spin $S = 0$, isospin $I = 0$ channel and is suppressed in systems with small proton fraction since neutron pairs have $I = 1$.

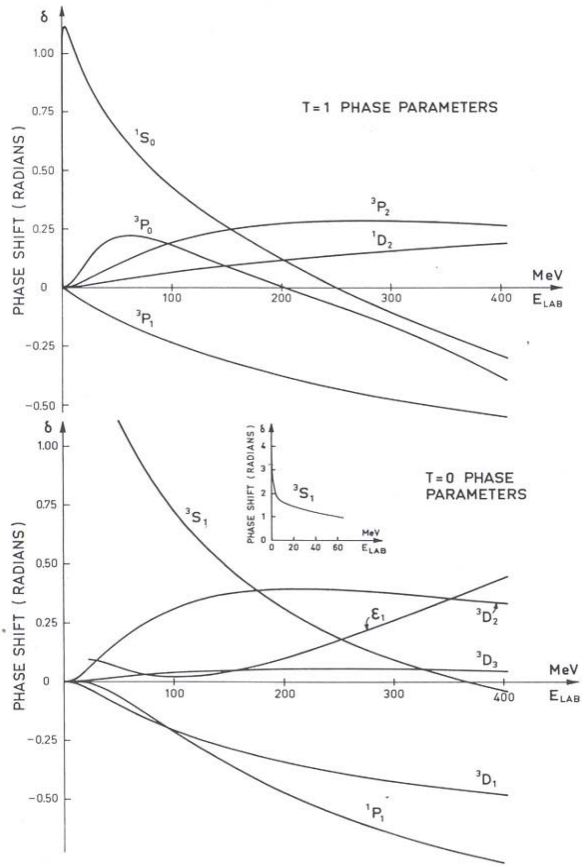

FIG.10. Behavior of the $S, P, D$ phase shifts with energy. The figure is from Bohr and Mottelson [39]. The isospin $I$ is labeled $T$ in their figure.

*D and higher order waves*

The interesting feature that the average $P$-wave phase shift $\bar{\delta}(^3P)$ is considerably reduced for triplet spin $S = 1$, isospin $I = 1$, $J = 0,1,2$ states no longer applies to spin $S = 1$, isospin $I = 0$ $D$-waves with $J = 1,2,3$. This feature can be seen from an analysis of Fig.10. In particular, the negative $^3D_1$ phase shift does cancel some of the positive $^3D_3$ and $^3D_2$ contributions, but the $^3D_2$ phase shift is larger than the $^3D_1$ phase shift for all energies up to ~200 $MeV$. Moreover the $^3D_2$ contribution to the continuum partition function is enhanced over the $^3D_1$ contribution because of the $2J + 1$ factor which is 5 for $J = 2$ and 3 for $J = 1$. The $^3D_3$ phase shift is small but also gets enhanced by the $2J + 1 = 7$ factor. In very small proton fraction systems, the dominant contribution for $D$-waves comes from the neutron-neutron isospin $I = 1$ component which is the singlet $S = 0$, $^1D_2$ phase shift which makes a positive contribution. The isospin $I = 0$ component of the neutron-proton system is associated with the triplet $S = 1$ spin state and ${}^3D_1, {}^3D_2, {}^3D_3$ waves. Unlike some of the $S$ and $P$-waves, none of the initially positive $D$-waves change sign. The ${}^3D_1$ and ${}^3S_1$ waves are also coupled by the tensor force. This feature will be ignored. Moreover, results based on a delta shell potential will be given. These results are shown in Fig.11.

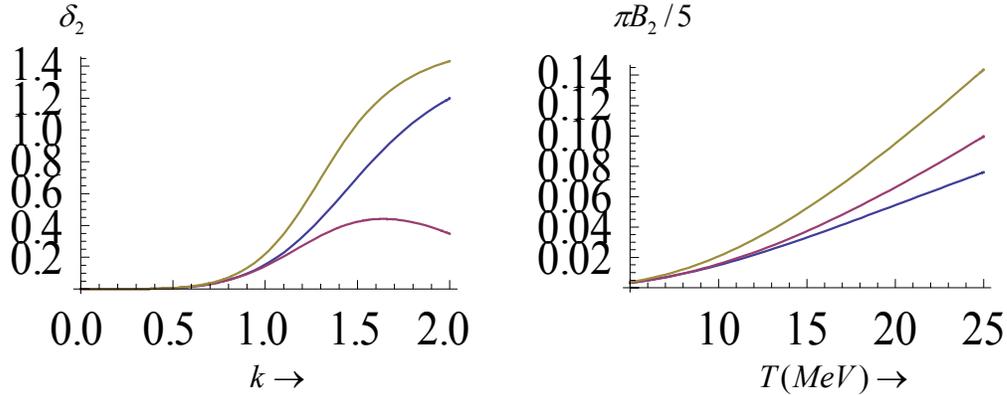

FIG.11. Left figure is the behavior of a $D$-wave phase shift with $k(fm^{-1})$. The plot was made using a delta shell attractive potential with strength $\chi = .6$ and radius $R = 2\,fm$. The lowest curve is the delta shell potential, the middle curve is the effective range approximation with scattering length $a_{sl}^2 = -.225$ and effective range $r_{eff}^2 = 4.02$. The upper curve has the effective range equal to zero. The parameters are chosen so that $\delta_2 = .43$ at an energy $100\,MeV$ which is near the value of $\delta_2({}^3D_2)$ shown in figure 9. The right figure shows the behavior of $\pi B_2 / 5$ with temperature for the same set of parameters. The curves are ordered as in the right figure.

It is also interesting to study the contribution of many partial waves to the second virial coefficient which occurs at high temperatures. The next subsection develops results for the special case of a hard sphere potential.

*Many partial waves and classical hard sphere behavior at very high energies*
The partial wave expansion for scattering from a hard sphere is given by [41]

$$\sigma = \frac{4\pi}{k^2} \sum_0^\infty (2l+1) \sin^2 \delta_l = \frac{4\pi}{k^2} \sum_0^\infty (2l+1) \frac{j_l^2(kc)}{j_l^2(kc) + \eta_l^2(kc)} \tag{56}$$

At high energies this cross section goes to the well known result $\sigma \to 2\pi c^2$ which is twice the classical value for scattering from a hard sphere. The extra two comes from diffraction scattering of the incident wave around the hard sphere. The quantity $d\delta_l / dk$ contains the same factor $j_l^2(kc) + \eta_l^2(kc)$ in the denominator that appears also in $\sigma$ for the hard sphere interaction. It is therefore worthwhile to see how the limit

$\sigma \to 2\pi c^2$ was obtained and find the scaling behavior of the same approximations to evaluate the sum

$$\sum_{l=0}^{\infty}(2l+1)\frac{d\delta_l}{dk} = -c\sum_{l=1}^{\infty}(2l+1)\frac{1}{(kc)^2(j_l^2(kc)+n_l^2(kc))} \quad (57)$$

The main approximations are lsted in Ref [41] and are: 1. the asymptotic values of $j_l(kc)$ and $n_l(kc)$ are used for all $kc$ that are larger than $l$, of the order of $l$, and smaller than $l$; 2. most of the contribution to the sum over $l$ comes from $l < kc - C(kc)^{1/2}, C \sim 1$. The following scaling behavior is found:

$$S_1 \equiv \sum_{l=0}^{x}(2l+1)\frac{j_l^2(x)}{(j_l^2(x)+n_l^2(x))} \to \frac{1}{2}x^2 \quad (58)$$

A similar analysis gives

$$S_2 \equiv \sum_{l=0}^{x}(2l+1)\frac{1}{(j_l^2(x)+n_l^2(x))} \to \frac{2}{3}x^4 \quad (59)$$

The scaling behavior of the last sum $S_2$ divide by $x^4$ is shown in the left part of Fig.12. The right part of this figure shows the scaling property of $S_1$ divided by $x^2$

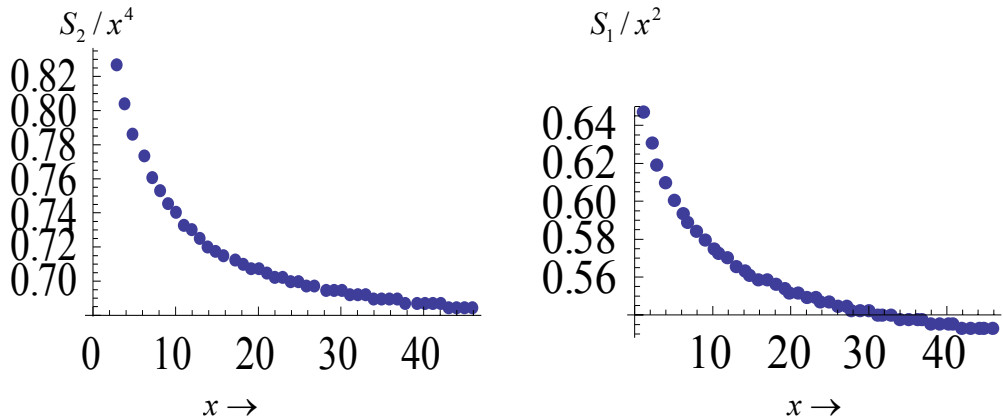

FIG.12. Scaling behavior of the sum $S_1$ and $S_2$. The left part of the figure shows $S_2$ divided by $x^4$ for $x$ from $5 \le x \le 50$. The limiting behavior of $S_2/x^4$ is $2/3$. The right part of the figure shows $S_1$ divided by $x^2$ for the same range of $x$ and has a limiting value $1/2$.

Thus, for $kc \gg 1$

$$\sum_{l=0}^{\infty}(2l+1)\frac{d\delta_l}{dk} = -c\frac{1}{(kc)^2}\frac{2}{3}(kc)^4 = -\frac{2}{3}k^2c^3 \tag{60}$$

and

$$\frac{1}{\pi}\int dk \sum_{l=0}^{\infty}(2l+1)\frac{d\delta_l}{dk}Exp(-bk^2) = -\frac{\sqrt{2}}{3}\pi\frac{c^3}{\lambda_T^3} \tag{61}$$

This result shows that the second virial coefficient when summed over many partial waves at high energies depends on the radius of the hard core as $c^3$ and thus the volume of a hard sphere. The result of Eq.(61) can be compared with the corresponding hard sphere $S-$wave result which involves ratio of the hard sphere radius to quantum thermal wavelength or $c/\lambda_T$. The van der Waals equation of state contains a hard core feature involving the volume $c^3$ which brings us to the discussion and comparison of the next subsection.

*Comparisons with van der Waals and Skyrme equations of state*
The second virial coefficient using classical statistical mechanics is [42].

$$b_2 = -\frac{2\pi}{\lambda_T^3}\int(1-\exp(-V(r)/k_BT))r^2 dr \tag{62}$$

For a square well potential with a hard core, the

$$b_2 = -\frac{2\pi}{\lambda_T^3}(\frac{c^3}{3} + (1-e^{-V_0/k_BT})\frac{R^3-c^3}{3}) \approx -\frac{2\pi}{\lambda_T^3}(\frac{c^3}{3} - \frac{V_0}{k_BT}\frac{R^3-c^3}{3}) \tag{63}$$

where the approximation in this equation is obtained from the limit $V_0/k_BT \ll 1$. The resulting equation of state $PV/Ak_BT = 1 - b_2\lambda_T^3 A/V$ gives

$$P = \frac{k_BT}{V/A}(1 + \frac{2\pi}{V/A}(\frac{c^3}{3} - \frac{V_0}{k_BT}\frac{R^3-c^3}{3}) \tag{64}$$

This last result rewritten and can be approximated as

$$(P + \frac{2\pi}{3}|V_0|(R^3-c^3)\frac{A^2}{V^2})(V - \frac{2\pi}{3}c^3A) \tag{65}$$

leading to an EOS of the van der Waals type: $(P + b_W(A/V)^2)(V - v_xA) = Ak_BT$. The $b_W(A/V)^2$ with $b_W > 0$ term is a reduction in pressure due to attractive interactions between pairs particles, atoms or molecules and $v_x$ is the excluded volume.

The Skyrme interaction has been used to model the thermodynamic features of nuclear matter. For example, recent work on the application of the Skyrme interaction to the liquid-gas phase transition can be found in Ref [43]. The Skyrme interaction contributes an interaction term to the energy per particle that is of the form $-a_0\rho + a_3\rho^{1+\sigma}$. We will consider the case $\sigma = 1$ for the purpose of a simplified illustration. In this case the equation of state at a high temperature reads

$$P = \rho k_B T - a_0 \rho^2 + 2a_3 \rho^3 \tag{66}$$

The coefficients $a_0, a_3$ are determined by requiring the Skyrme interaction to give the binding energy per particle of nuclear matter to be $E_{nm}(=-16 MeV)$ at nuclear matter density $\rho_0 (=.17/fm^3)$ at $T = 0$. The Fermi energy $E_F = 35 MeV$.
For general $\sigma$

$$a_0 = \frac{\frac{1+\sigma}{\sigma}|E_{nm}| + \frac{1+3\sigma}{3\sigma}E_K}{\rho_0}, \quad a_3 = \frac{\frac{1}{\sigma}|E_{nm}| + \frac{1}{3\sigma}E_K}{\rho_0^{1+\sigma}}, \quad E_K = \frac{3}{5}E_F \tag{67}$$

In both the Skymre and the van der Waals EOS an attractive term only appears in the second virial coefficient or $\rho^2$ term. The repulsive parts of the interaction are treated as a hard core excluded volume subtraction in the van der Waals EOS. In the Skyrme EOS, the repulsion appears in a higher power density dependent term which is $\rho^{2+\sigma}$. While both these EOS are simplified descriptions, they both offer a qualitative description of phase transitions. Higher order attractive virial terms arise when larger clusters than size 2 are included. Attempts have been made to obtain the third virial coefficient for the nuclear EOS [44]. The next subsection gives a discussion of higher order virial expansion.

*Higher order clusters in a two component system*
For a two component system with initially $Z$ protons and $N$ neutrons, the formation of clusters of type $N_{ij}$ with $i$ protons and $j$ neutrons has the following equations.
 The two component Saha equation is $N_{ij} = a_{ij}(N_p)^i(N_n)^j$. The law of partial pressures reads:

$$\frac{PV}{k_B T} = \sum_{i,j=0,1,2,...} N_{ij} \tag{68}$$

with $N_{10} = N_P$ and $N_{01} = N_n$. Consider clusters up to mass 4. Two constraint equations exist: $N_p = yA - N_d - N_t - 2N_{He^3} - 2N_\alpha$ and $N_n = (1-y)A - N_d - 2N_t - N_{He^3} - 2N_\alpha$. The $A = Z + N, y = Z/A, (1-y) = N/A$. The resulting EOS up to $A^4$ terms is

$$\frac{PV}{k_BT} = A - a_d y(1-y)A^2 - y(1-y)A^3\{-a_d^2 + 2a_t(1-y) + 2a_{He^3}y\} +$$

$$y(1-y)A^4\{-a_d^3(1+y(1-y)) + 3a_d a_t(1-y)(1+y) + 3a_d a_{He^3}y(2-y) - 3a_\alpha y(1-y)\} \quad (69)$$

The two component EOS at $y = 1/2$ goes into a one component EOS with $N_j = a_j(N_1)^j$ under the substitutions $a_d \to 2^2 a_2, a_t + a_{He^3} \to 2^3 a_3, a_\alpha \to 2^4 a_4$. Excluding spin $J$ and excited states, the $a_j = j^{3/2} Exp[E_b(j)/k_BT](\lambda_T^3/V)^{j-1}$. The $a_i$ coefficients appear in the law of mass action equations: $N_d = a_d N_p N_n$, $N_t = a_t N_p N_n^2$, $N_{He^3} = a_{He^3} N_n N_p^2$, $N_\alpha = a_\alpha N_p^2 N_n^2$. The one component EOS is

$$\frac{PV}{k_BT} = A - a_2 A^2 + (4a_2^2 - 2a_3)A^3 + (-20a_2^3 + 18a_2 a_3 - 3a_4)A^4 \quad (70)$$

The structure of this equation is the same as Eq.(8) when $a_1 = 1$ and with $a_2 = x_2/x_1^2$, $a_3 = x_3/x_1^3$, $a_4 = x_4/x_1^4$. One of the difficulties in evaluating higher order virial coefficients is the contribution of all the various continuum contributions. These continuum correlations can cancel in part the bound state contributions as they do for the second virial coefficient. Thus only some qualitative remarks can be made.

Because of its large binding energy and several excited states, the alpha particle makes significant contributions to the thermal behavior of the system. The number of alpha particles is given by

$$N_\alpha = 4^{3/2} \frac{1}{2^4} \frac{(\lambda_T^3)^3}{V^3} \sum_i (2J_i + 1) Exp[\frac{|E_i|}{k_BT}](N_p)^2(N_n)^2 \quad (71)$$

The sum that appears is over the ground and excited states in the $\alpha$ particle. The $\alpha$ particle makes a contribution to the fourth virial coefficient, the deuteron appears in the second, third, fourth,…coefficients. The triton and $He^3$ first appear in the third coefficient and also in subsequent terms. For example the fourth coefficient in Eq.(69) has terms involving $a_d^3, a_d a_t, a_d a_{He^3}, a_\alpha$. The ratio of ground state $\alpha$ particles $N_{\alpha.gs}$ to deuterons $d$ is

$$N_{\alpha.gs}/N_d = (N_p \lambda_T^3/V)(N_n \lambda_T^3/V)(\sqrt{2}/6) Exp\{(E_b(\alpha) - E_b(d))/k_BT\} =$$

$$(N_d \lambda_T^3/V)(1/9) Exp\{(E_b(\alpha) - 2E_b(d))/k_BT\}. \quad (72)$$

The cluster yields decrease rapidly with $A$ when the nucleon density $\rho_N$ and $\lambda_T^3$ satisfy $(\rho_N \lambda_T^3/4) exp(e_b/k_BT) << 1$ for $y = 1/2$. The $e_b$ is the binding energy per particle. Moreover a signal for the presence of very large clusters is the formation of the liquid-gas phase transition which occurs at $k_B T \sim 9 MeV$ for $\rho_N \equiv \rho_{LG} \sim .075$ nucleons/fm$^3$. At temperatures $k_B T \approx e_b = 8 MeV$, the Boltzmann factor in binding energy plays an important role. In fact, because of the large binding effect of the $\alpha$ particle compared to the deuteron ($E_b(\alpha) - 2E_b(d) \approx 24 MeV$), the $\alpha$ particle can dominate at low $T$ [18,19] because of the Boltzmann factor in binding energy. The proton fraction $y$ also plays a significant role in systems with small $y$ since the number of nuclei of the type $_Z A_N$ is proportional to $y^Z (1-y)^N$. Thus an $\alpha$ particle will be more suppressed compared to a deuteron when $y \approx 0$ since $N_\alpha \sim y^2, N_d \sim y$. For mass 3 clusters, which appear in the third virial term, the number of $He^3$ is reduced by $y/(1-y)$ compared to the number of tritium $t$. At small $y$ the curly bracket term in Eq.(69) for the fourth virial coefficient is $\{-a_d^3 (1+y) + 3a_d a_t + 6y a_d a_{He^3} - 3y a_\alpha\}$. For systems with small $y$, the quantity $y(\rho_N \lambda_T^3 / 4) \exp(e_b / k_B T)$ will determine the rate of fall off clusters yields with one more proton and mass number one unit higher. Very small $y$ are present in a neutron star [45]. Low concentrations of proton and electrons are necessary to Pauli block neutron $\beta$ decay. The equilibrium is reached through $n \to p + e^- + \bar{\nu}_e$ and $p + e^- \to n + \nu_e$ with the neutrinos escaping. The inner crust of neutron stars is also an example where dilute Fermi gases are present [46].

*Isospin conserving case*
In the limit the spin singlet $l = 0$ channel scattering lengths $a_{ij,S} = a_{nn,0} = a_{np,0} = a_{pp,0}$ are the same and the triplet $a_{ij,1}$ and singlet $a_{ij,0}$ are such that $b << a_{ij,1}$ and $a_{ij,0}$ then $\hat{b}_{2,\text{int}}$ is

$$\hat{b}_{2,\text{int}} = \frac{1}{4} 2^{3/2} \lambda_T^3 \left( 3y(1-y) \exp\frac{E_B}{k_B T} + (1-2y)^2 + 3y(1-y) \frac{\sqrt{2}}{\pi} \frac{\lambda_T}{|a_{np,1}|} - (1 - y(1-y)) \frac{\sqrt{2}}{\pi} \frac{\lambda_T}{|a_{np,0}|} \right)$$

(73)

The above formulae does not include $r_0$ corrections to keep the expression simple for the moment. In the universal thermodynamic limit with $E_B \to 0$ and all scattering lengths $a_{ij} \to \infty$ and now including effective range corrections, the $\hat{b}_2 / \lambda_T^3$

$$\hat{b}_2 / \lambda_T^3 = \{\frac{3}{2^{7/2}} - \frac{2^{3/2}}{4}(\frac{1}{6\sqrt{\pi}}\frac{ro}{\sqrt{b}})\} - y(1-y)\{\frac{3}{2^{5/2}} - \frac{2^{3/2} 2}{4}(\frac{1}{6\sqrt{\pi}}\frac{ro}{\sqrt{b}})\}$$
(74)

The $\hat{b}_2$ includes anti-symmetry exchange correlations and the above formula is for an isospin symmetric case with all effective ranges taken to be the same including $a_{np,0}$ & $a_{np,1}$ for simplicity.

**Conclusions.** A finite temperature two component model of strongly correlated protons and neutrons each with two spin states and underlying isospin symmetries was discussed. The model is an extension of the one component fermionic models in both atomic systems and in nuclear physics where pure neutron systems are considered. Features associated with Feshbach resonances and bound states can be studied by tuning on the proton fraction $y$ and varying the temperature $T$. In atomic systems the tuning is done in a controlled way by a magnetic field which can change the scattering length across infinity. In the nuclear case the bound state is the spin $S=1$, isospin $I=0$ state of the $np$ system, which is the loosely bound deuteron. The resonant like virtual structures arise in the $S=0$, $I=1$ channels. The $nn$ and $np$ $S=0$, $I=1$ channels have very large scattering lengths and approximate the infinite limit. The mixture with varied neutron/proton asymmetry has both positive and negative scattering lengths and is therefore different from the one component atomic case tuned by the magnetic field. When the scattering lengths $a_{sl}$ are large compared to interparticle spacings and range of interparticle forces, a regime called the unitary limit is reached and this limit was studied. A simplified interaction between nucleons was used which is an attractive square well potential with a short range hard core repulsion. Properties of this potential were then related to experimental results for nucleon-nucleon effective ranges $r_0$ and $a_{sl}$ in spin singlet and triplet states. Analytic results (Eq.(41-43)) were developed for the Beth Uhlenbeck continuum correlation integral which very accurately describes exact results based on the potential models discussed- see Fig.5-6. These analytic results were then used to study various features such as the interaction energy, the EOS, the interaction entropy and the isothermal compressibility. In the unitary limit of infinite scattering length, the ratio of the effective range to quantum volume, or $r_0/\lambda_T$, appears as a limiting scale in these various thermodynamic quantites. A rescaled interaction energy $\hat{\varepsilon}_{int}$ was shown to have corrections which are linear in $r_0/\lambda_T$, with no quadratic, cubic terms (or even fourth order terms). The entropy has no linear, quadratic or fourth order terms, and starts with a cubic term in this ratio. The compressibility to the ideal gas law has both a linear and a cubic term. The role of bound states in thermodynamic quantities was also considered. The deuteron bound state produce a large departure in the interaction energy from the unitary limit and giving rise to a pronounced $T$ dependence at low $T$ as already shown in Ref. [27]. Higher order clusters are also important, and in particular the $\alpha$ particle, can dominate [18] in a virial expansion in certain regions of proton fraction $y$ and temperature $T$. Comparisons were also made with hard sphere Bose and Fermi gases which are limiting cases of the potential used. The contribution of many partial waves to the equation of state for a hard sphere gas was shown to result in the appearance of the hard core volume term similar to the van der Waals case. This result is a consequence of a scaling law similar to that encountered for hard sphere scattering and an associated enhancement of the cross section by a factor of two from diffraction around the hard sphere. Also considered was another limiting situation which is a delta-surface interaction. Results for the delta surface interaction were shown to closely approximate a square well interaction for the thermodynamic quantities considered. The phase shifts between the two potentials differ considerably at large $k$, but the Boltzmann factor in energy suppresses the high $k$ part of the phase shift were they differ. The delta-shell interaction is easier is to work with simple closed form expressions for phase shifts in various angular states.

*Appendix A. Hard sphere Bose and Fermi gas properties,*
An attractive potential with a hard core reduces to a pure hard sphere of radius $r = c$ when $V_0 \to 0$ and $R \to c$. The hard sphere limit will be used also to compare with results for a hard sphere spinless Bose gas and a hard sphere spinless Fermi gas given in Huang [31] or Pathria [42]. For a hard sphere of radius $c$,

$$\tan \delta_l = j_l(kc) / \eta_l(kc) \tag{A.1}$$

and

$$d\delta_l / dk = -c/(kc)^2 \{(j_l(kc))^2 + (\eta_l(kc))^2\}. \tag{A.2}$$

The phase shifts are decreasing with $k$ since the interaction is a repulsive hard core. For an $S$-wave $\delta_0 = -kc$ and $d\delta_0 / dk = -c$. For a $P$-wave $\delta_1 = -kc + \arctan(kc)$ and $d\delta_1 / dk = -c(kc)^2 /(1 + (kc)^2)$. A $D$-wave has $\delta_2 = -kc + \arctan[3kc/(3-(kc)^2)]$ and $d\delta_2 / dk = -c(kc)^4 /(9 + 3(kc)^2 + (kc)^4)$. Higher partial waves can be obtained from recurrence relations. Writing $j_l = a_l \sin(x) - b_l \cos(x)$, $\eta_l = -a_l \cos(x) - b_l \sin(x)$, the $j_l^2 + \eta_l^2 = a_l^2 + b_l^2$. The recurrence relation $j_{l+1}(x) = ((2l+1)/x) j_l(x) - j_{l-1}(x)$ leads to $a_{l+1}(x) = ((2l+1)/x)a_l(x) - a_{l-1}(x)$, $b_{l+1}(x) = ((2l+1)/x)b_l(x) - b_{l-1}(x)$. The $a_0 = 1/x$, $a_1 = 1/x^2$, $b_0 = 0$ and $b_1 = 1/x$. For example $a_3 = 15/x^4 - 6/x^2$, $b_3 = 15/x^3 - 1/x$, giving $d\delta_3 / dk = -c(kc)^6 / 225 + 45(kc)^2 + 6(kc)^4 + +(kc)^6)$

From the above specific results, the structure of $d\delta_l / dk$ is

$$\frac{d\delta_l}{dk} = -c \frac{x^{2l}}{x^{2l} + d_{2l-2} x^{2l-2} + .... + ((2l-1)!!)^2} \tag{A.3}$$

The denominator is a polynomial in even powers of $x$ from $x^0$ to $x^{2l}$. The $(2(l-1)!!)^2$ coefficient of $x^0$ serves as a cut off for high $l > kc$ terms.
  The integrals over the rate of change of the phase shift are:

$$2^{3/2} \frac{1}{\pi} \int \frac{d\delta_0}{dk} \exp(-\hbar^2 k^2 / mk_B T) dk = -2 \frac{c}{\lambda_T}$$

$$2^{3/2} \frac{1}{\pi} \int \frac{d\delta_1}{dk} \exp(-ak^2) dk = -c^3 (\frac{\sqrt{\pi}}{2\sqrt{a}c^2} - \frac{\exp(\frac{a}{c^2})\pi Erfc(\frac{\sqrt{a}}{c})}{2c^3}) \sim -6\pi(\frac{c}{\lambda_T})^3 + 18\pi^2(\frac{c}{\lambda_T})^5$$

$$2^{3/2}\frac{1}{\pi}\int\frac{d\delta_2}{dk}\exp(-ak^2)dk \sim -\frac{10}{3}\pi^2(\frac{c}{\lambda_T})^5 \tag{A.4}$$

A spinless hard sphere Bose gas [31,42] would then have an equation of state

$$\frac{1}{A}\frac{PV}{k_BT}=1+\left(-\frac{1}{2^{5/2}}+2\frac{c}{\lambda_T}+\frac{10}{3}\pi^2(\frac{c}{\lambda_T})^5\right)\frac{A}{V}\lambda_T^3 \tag{A.5}$$

with the $c/\lambda_T$ term the $l=0$ contribution and the $(c/\lambda_T)^3$ term the $l=2$ part.

A "spinless" hard sphere Fermi gas considered in ref. [31,42], subject to antisymmetrization effects from space considerations only, would then have an equation of state from $l=1, P-$wave contributions that is

$$\frac{1}{A}\frac{PV}{k_BT}=1+\left(\frac{1}{2^{5/2}}+6\pi(\frac{c}{\lambda_T})^3-18\pi^2(\frac{c}{\lambda_T})^5\right)\frac{A}{V}\lambda_T^3 \tag{A.6}$$

The $l=1, P-$wave terms shown are the expanded expression for the exact $P-$wave expression in Eq.(A.4).

Spin ½ fermions of the same type can interact with $l=0,2$ contributions when the total spin $S=0$ and $l=1,3$ when the total spin is $S=1$. A system of protons and neutrons has terms that arise from fermions of the same type - *pp* & *nn* -and fermions that are different - *np*. The *np* system in the isospin $I=1$ channel with their spins coupled to spin $S=1$ has $l=1,3$ contributions and when their spins are coupled to spin $S=0$ has $l=0,2$ contributions. The *np* system in the isospin $I=0$ channel with spins coupled to spin $S=1$ has $l=0,2$ contributions and with spins coupled to spin $S=0$ has $l=1,3$ contributions The antisymmetrization factor is now changed from $1/2^{5/2} \to 1/2^{7/2}$ arising from an additional $1/(2S+1)$ spin term.

## *Appendix B. Surface-Delta or Delta-Shell Potential*

Another special case of a square well potential that can be considered is a surface delta interaction with shape $V(r)=-\chi\delta(r-R)$. The surface-delta interaction is considered in great detail in Gottfried [36] and is briefly included here because it leads to simple expressions which are useful in understanding various features. Results from the surface delta can also be compared with a square well potential interaction. The phase shifts can be shown to be given by a simple closed form expression that is

$$\tan\delta_l = \frac{(\chi R)(kR)j_l^2(kR)}{1+(\chi R)(kR)j_l(kR)\eta_l(kR)} \tag{B.1}$$

In the limit $g \equiv \chi R \to \infty$, the $\tan \delta_l$ is the same as that of a hard sphere. The $S$-wave phase shift is determined from

$$k \cot \delta_0 = \frac{k}{-g \sin(kR)\cos(kR)/kR} - k \cot(kR) \tag{B.2}$$

For small $ka$,

$$k \cot \delta_0 = -(g-1)/Rg + 2((1+g)R/3g) \cdot k^2/2 = -\frac{1}{a_{sl}} + \frac{1}{2} r_o k^2 \tag{B.3}$$

which is an effective range expansion. For arbitrary $l$

$$k^{2l+1} \cot \delta_l = -\frac{((2l+1)!!)^2}{(2l+1)R^{2l+1}} \left(\frac{g-(2l+1)}{g}\right) + \frac{1}{2} k^2 \left( \frac{R^2}{R^{2l+1}} \frac{2((2l+1)!!)^2}{(2l+3)(2l-1)} \frac{((2l-1)-g)}{g} \right) =$$

$$= -\frac{1}{a_{al}^l} + \frac{1}{2} r_{eff}^l k^2 \tag{B.4}$$

An $S$-wave bound state appears when $g \geq 1$ and has zero binding energy when $g = 1$. At most a single $S$-wave bound state occurs.
The scattering amplitude is given

$$Exp[i\delta_l] \sin \delta_l = \frac{\tan \delta_l}{1 - i \tan \delta_l} = \frac{1 + kR\gamma Rj_l^2(kR)}{1 - ikR\gamma Rj_l(kR)h_l^1(kR)} \tag{B.5}$$

with $h_l^1 = j_l + i\eta_l$. Defining $\eta = kR$ the zeroes of the denominator

$$D[g, \eta] = 1 - i\eta g j_l(\eta) h_l^1(\eta) \tag{B.6}$$

in the complex plane $\eta \to \zeta = \xi + i\eta$ determine the bound states and resonances. For small $\zeta$, $j_l = \zeta^l/(2l+1)!!$, $\eta_l = -(2l-1)!!/\zeta^{l+1}$ and thus $j_l h_l^1 = -i/(2l+1)\zeta$ leading to

$$D[g, \zeta \to 0] = 1 - \frac{g}{2l+1} \tag{B.7}$$

Thus a zero energy bound state with angular momentum $l$ appears when $D[g, \zeta \to 0] \to 0$, which gives the connection $g = 2l+1$. When $g > 1$, the $S$-wave bound state energy $E_b = -\hbar^2 \eta^2/2\mu R^2$ is obtained by solving the eigenvalue equation

$D_0[g, \zeta \to i\eta] = 1 + (g/2\eta)(Exp[-2\eta] - 1) = 0$. For $g = 1 + \varepsilon$ with $\varepsilon \ll 1$, the $\eta_b = \varepsilon$ and the $E_b = -\hbar^2 \varepsilon^2 / 2\mu R^2$. The bound states are poles of the scattering amplitude on the positive imaginary axis. A set of poles in the lower half of the complex $\zeta$ plane with $\xi \neq 0$ are also present which can be determined by solving $D_l[g, \xi + i\eta] = 0$.

The factor $\sin^2 \delta_0 = \tan^2 \delta_0 / (1 + \tan^2 \delta_0)$ that appears in the $S$-wave contribution to the total cross section is

$$\sin^2 \delta_0 = \frac{g^2 \sin^4 \xi}{(\xi - (g/2) \cdot \sin 2\xi)^2 + g^2 \sin^4 \xi} \tag{B.8}$$

The condition $\sin^2 \delta_0 = 1$ is at a finite set of points where $(\xi - (g/2) \cdot \sin 2\xi)^2 = 0$ or $2\xi = g \sin 2\xi$. At these points the cross section goes to its unitary maximum $4\pi/k^2$. The behavior of $\sin^2 \delta_0$ and that of a hard sphere, where $\sin^2 \delta_0 = \sin^2 \xi$, is shown in Fig.B.1

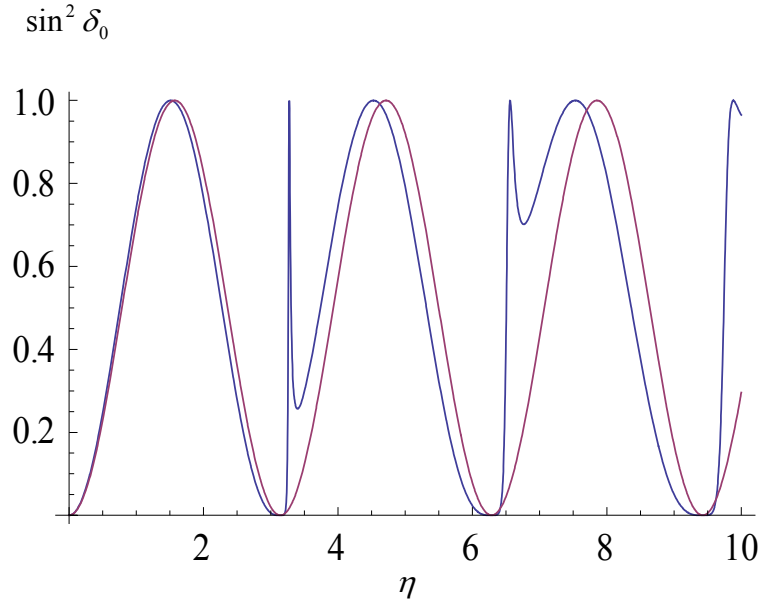

Fig.B.1. Surface-delta potential versus hard shere potential as a function of the dimensionless variable $\eta = kR$. The hard sphere radius and surface-delta radius are the same. The hard sphere $\delta_0 = -kR = -\xi$. The $g = 25$. The surface delta curve is very similar to the hard sphere except at $\sim \pi, 2\pi,...$ where the surface delta has extra spikes which correspond to sharp resonances as already discussed. The hard sphere limit is reached when $g \to \infty$. Other maxima are at $\pi/2, 3\pi/2,...$ for a hard sphere and near these points for the surface-delta. These points are broad maxima. Once narrow resonances become broad such that the width becomes comparable to the spacing the resonances peaks will disappear in a cross section.

The delta-surface interaction has simple expressions for the scattering length $a_{sl}$ and the effective range $r_0$ given by $a_{sl} = Rg/(g-1)$ and $r_0 = 2(1+g)R/3g$ for $S$-waves. The scatttering length can be either positive or negative depending on $g-1$.

Fig.B.2 compares the $P$-wave phase shift for a repulsive the delta shell potential with the phase shift calculated with a repulsive square well step potential.

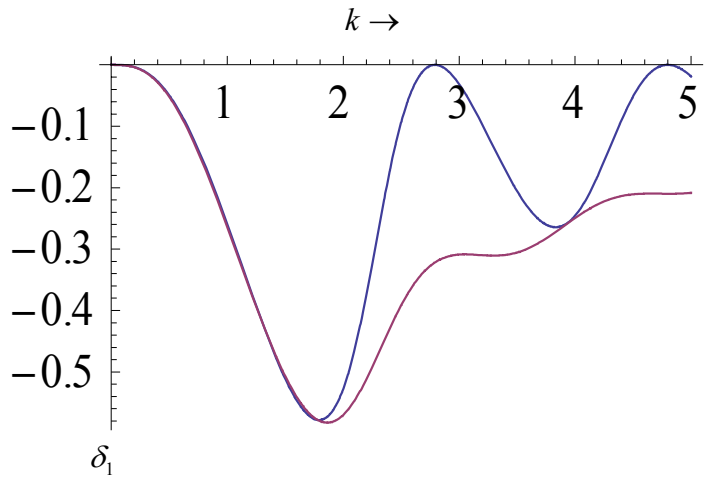

FIG.B.2. The $P$-wave phase shift in radians calculated with a repulsive delta shell potential and repulsive square well. The delta shell potential is the upper curve The delta shell potential has strength $\chi = -1$, radius $a = 1.61\,fm$, strength $g = \chi a = -1.61$ while the square well potential has $|\alpha_0| = 1/\,fm$ and radius $R = 2\,fm$. The delta shell potential parameters were chosen to fit the square well results from $0 \le k \le 1.5\,fm^{-1}$.

Fig.3 is a comparison of the calculation of the continuum integral $B_c(P-\text{wave})$ for a delta shell repulsive potential, a repulsive step potential and a scattering length, effective range approximation. The figure shows that $B_c(P-\text{wave})$ calculated in the three different ways are very close to each other to the point of almost being indistinguishable. This suggests using a delta range potential for calculations of various quantities that appear in this paper.

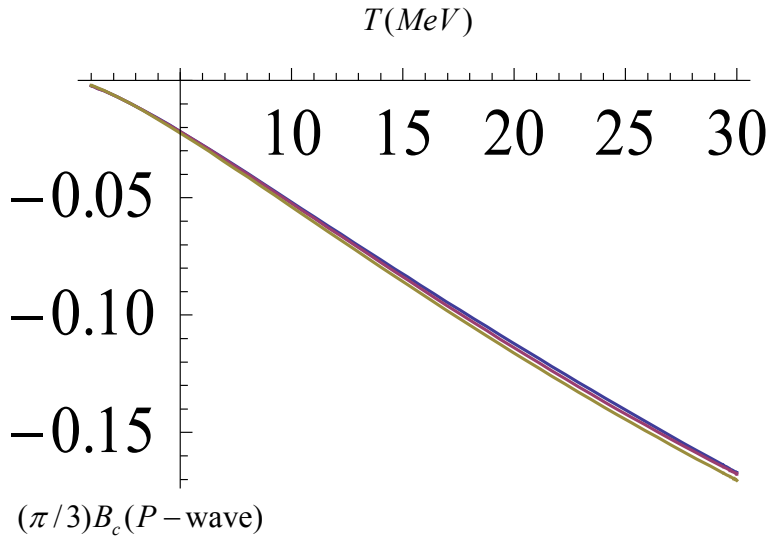

Fig.B.3. Contributions to $B_c(P-\text{wave})$. The three contributions shown are from a delta shell potential (upper curve), a repulsive step potential (middle curve which is within the thickness of the upper line and can't be distinquished from it) and an effective range approximation (lowest curve). The parameters of the delta shell potential and the repulsive step potential are the same as in Fig.B.2. The effective range parameters are those of the delta shell potential for a $P-\text{wave}$ and are $a_{sl}=.486, r_{eff}=-3.625$. The effective range results are very good even at high temperatures because of the Boltzmann factor suppression of high energy contribution.

**Acknowledgments.** This work supported by Department of Energy under Grant DE-FG02ER-40987.